\documentclass[aps,prb,twocolumn,twoside,letterpaper,showpacs,amssymb]{revtex4}
\usepackage{graphicx}
\usepackage{color}
\usepackage{amsmath}
\usepackage{enumerate}
\def\T{T}
\def\cN{{\cal N}}
\def\vv{{\bf v}}
\def\vp{{\bf p}}
\def\vz{{\bf z}}
\def\vR{{\bf R}}

\newcommand{\D}{\mathfrak{D}}
\newcommand{\G}{\mathfrak{G}}

\newcommand{\hG}{\hat{\mathfrak{G}}}
\newcommand{\grad}{\text{\boldmath$\nabla$}}
\newcommand{\Area}{{\cal A}}

\newcommand{\Del}{{\mbox{\footnotesize $\Delta$}}}
\begin{document}
\title{Heat Transport Through Josephson Point Contacts}
\author{Erhai Zhao, Tomas L$\mathrm{\ddot{o}}$fwander, and J. A. Sauls}
\affiliation{Department of Physics \& Astronomy, Northwestern University, Evanston, IL 60208}
\begin{abstract}
We present a comprehensive study of heat transport through small
superconducting point contacts. The heat current for a temperature biased weak
link is computed as a function of temperature and barrier transparency of the
junction. The transport of thermal energy is controlled by the quasiparticle
transmission probability for the point contact that couples the
superconducting leads. We derive this transmission probability and results for
the heat current starting from nonequilibrium transport equations and
interface boundary conditions for the Keldysh propagators in quasiclassical
approximation. We discuss the thermal conductance for both clean and
dirty superconducting leads, as well as aspects of the nonlinear current
response. We show that the transmission probability for continuum
quasiparticle states is both energy- and phase-dependent, and controlled by an
interface Andreev bound state below the continuum. For high transparency
barriers the formation of a low-energy bound state, when the phase is tuned to
$\phi=\pi$, leads to a reduction of the heat current relative to that for
$\phi=0$. For low-transparency barriers, a shallow Andreev bound state
just below the continuum edge is connected with resonant transmission of
quasiparticles for energies just above the gap edge, and leads to enhanced heat
conductance as the temperature is lowered below the superconducting transition.
\end{abstract}
\date{\today}
\maketitle
\section{Introduction}

In recent years theoretical and experimental investigation of quantum transport
through small superconducting hetero-structures has increased significantly
(see for example Refs.~\onlinecite{agr03,bel99} and references therein).
These structures vary from length scales of the order of a micrometer down to
the atomic scale. Experimental progress is driven in part by the development
of atomic-scale break junctions, sub-micron- to nano-scale lithography and
the fabrication of novel multi-terminal mesoscopic structures,
as well as experimental techniques and methods of measuring the local temperatures
in small mesoscopic structures.\cite{aum99}

Considerable effort has gone into the study of charge transport in
mesoscopic devices, both the non-dissipative superconducting response as
well as the dissipative current response in voltage-biased
junctions.\cite{agr03} Much less is known about the electronic heat
transport in temperature-biased junctions. Here we provide a detailed
report of theoretical results on thermal transport in small
temperature- and phase-biased Josephson junctions and weak-links as a
function of barrier transparency, $D$, temperature, $T$, and phase bias,
$\phi$. We provide a derivation of the transport properties of
superconducting point contacts starting from the Keldysh formulation of
the nonequilibrium transport equations for the quasiclassical Green
functions, and extend our brief report in Ref.~\onlinecite{zha03c} of
the thermal conductance (linear response) to the non-linear current
response. We use the Ricatti formulation of the transport equations and
follow closely the notation in Refs. \onlinecite{esc99,esc00}.

Small Josephson junctions are ideal systems for investigating quantum effects
on transport under well defined non-equilibrium situations. Here we focus on
non-equilibrium transport induced by a temperature bias $\Del T$ across the
junction. The thermal current is shown to depend on the phase difference, $\phi$, across
the junction. To quasiclassical accuracy, i.e. to leading order in the small
parameters $\textsf{s}=(T_c/\epsilon_f, 1/k_f\xi_0, ...)$, the thermo-power is
generally negligible, and in this limit there is no temperature-induced
voltage across the junction. The phase remains constant, and thus the
temperature-biased junction is in a stationary state. But, in general the
thermo-power is non-zero, so to be more precise we estimate the relative
magnitude of the dissipative charge current to thermal current induced by the
temperature bias to be of order $J_{\text{\tiny e}}\sim
\textsf{s}(e/k_B)\,J_{\text{\tiny q}}$ where
$\textsf{s}\sim(T_c/\epsilon_f)\ll 1$ is the relevant particle-hole asymmetry
parameter. In a closed circuit particle-hole asymmetry leads
to a temperature-induced voltage of order $eV\sim k_B\Del T(T_c/\epsilon_f)$.
Thus, for measurements of the heat current on a timescale
$t\ll\tau\sim(\pi/\Del T)(\epsilon_f/T_c)$, the phase is constant and we may
consider the thermal current response in a stationary situation.
This is in contrast to the dissipative charge current for a voltage-biased
junction, where the voltage is generally so large that the a.c. Josephson effect
develops. The charge current is then composed of a phase-independent d.c. current
and the a.c. Josephson current which oscillates with frequencies $n\omega_J$,
where $\omega_J=2eV/\hbar$ is the Josephson frequency and $n$ is an integer
number.\cite{jos62,arn87} Thus, the temperature-biased junction opens up the
possibility of studying quasi-stationary, phase-dependent non-equilibrium
transport processes.

At a junction between two superconductors, there are in general bound states
in the quantum well formed between the superconducting order parameters on the
two sides.\cite{and65} In a clean system, without normal backscattering at the
junction ($D=1$), the formation of Andreev bound states (ABS) can be
understood in simple terms by considering a quasiclassical path: a closed path
involving one electron-leg and one hole-leg, the two legs connected by Andreev
reflections at the two sides. Since the superconducting phase is encoded in the
coherence factor during Andreev reflection, the resulting bound state energies
depend on the phase. For contacts in which the distance between the electrodes
is small compared to the coherence length, there is only one pair of bound
states, with energies that disperse with the phase as
$\epsilon_B=\pm\Delta\sqrt{1-D\sin^2(\phi/2)}$, where $0<D\leq 1$ is the
transparency of the interface barrier. Since an electric charge $2e$ is
transferred during Andreev reflection, the bound states participate in charge
transport. In fact the phase dispersion determines the electric
supercurrent-phase dependence through the relation
$I=(2e/\hbar)\partial\epsilon_B/\partial\phi$. The continuum quasiparticle
states do not contribute to the charge current; the bound states
determine the current-phase characteristics.

The situation is quite different when we consider heat transport. As is well known,
Andreev reflection\cite{and64} results in strong suppression of the heat current
through a normal-superconducting (NS) interface; an electron and a retro-reflected
hole have equal, but oppositely directed, probability currents. Thus, only a fraction
of the continuum quasiparticle states above the gap which are not retro-reflected
contribute to the heat current across an NS interface. For two superconductors coupled
by a point-contact weaklink (ScS) the local excitation spectrum near the contact plays
a central role in regulating the heat current through the interface. Although the
continuum states above the gap carry the heat, the transmission probability of these
excitations is strongly influenced by an interface Andreev bound state (ABS) at the
point contact. Since the binding energy and spectral weight of the ABS are controlled
by the relative phase of the two superconductors, the transmission probability and the
resulting heat current are also phase dependent. For high transmission barriers the
reduction of the continuum density of states by the formation of the ABS for $\phi\ne
0$ leads to a suppression of the heat current compared to the case with $\phi=0$. For
lower transmission barriers $D\lesssim 0.4$ and $\phi\neq 0$, resonant transmission of
quasiparticles with energies just above the gap leads to an increase in the heat
transport relative to the normal state at $T_c$, over a wide temperature region below
$T_c$.\cite{zha03c} In this paper we investigate this resonance effect and the heat
current in detail, including the dependence of the heat transport on the relative
phase, the barrier transparency, the barrier model, impurity scattering, as well as
the sensitivity of the heat transport to a finite temperature bias $\Del T$, i.e.
beyond the linear response limit for the conductance.

The paper is organized as follows. In Sec.~\ref{sec:model} we describe our model for
small Josephson junctions and point-contact weak links. In Sec.~\ref{sec:ballistic}
we derive the heat current for ballistic leads. We present
the results for the linear response in Sec.~\ref{sec:linear}, and the non-linear
response in Sec.~\ref{sec:nonlinear}. We also discuss the tunnel limit $D\ll 1$ in
some detail, and examine the singularity that is encountered within the tunnel
Hamiltonian (tH) method that was previously used to calculate the heat conductance of
an SIS tunnel junction.\cite{mak65,gut97,gut98} We include a discussion of the
heat current based on the tH method in the Appendix. Finally, the effect of disorder in
the superconducting leads is examined in Sec.~\ref{sec:diffusive} for superconductors
in the diffusive limit.

\section{\label{sec:model}Model}
\begin{figure}
\includegraphics[height=2in]{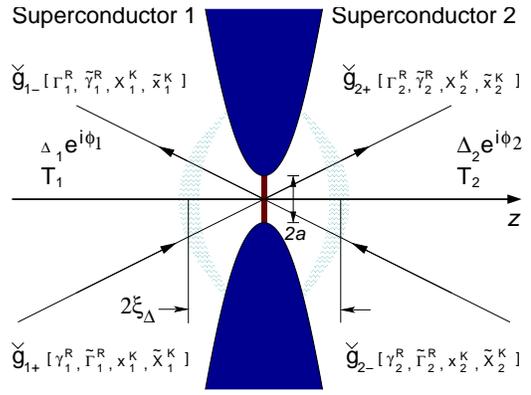}
\caption{Geometry of the temperature-biased Josephson contact of
      radius $a$. There is a potential barrier at $z=0$ with
      transparency $D(\vp_f)$, which connects trajectories with
      positive and negative projection of the Fermi momentum on
      the $\vz$-axis. The labels indicate which Ricatti amplitudes
      are computed along each trajectory. The shaded boundaries define the region
      ($\sim\xi_{\text{\tiny $\Delta$}}\gg a$) where
      superconductivity and the excitation spectrum are strongly
      modified for $\phi_1\ne\phi_2$.}
\label{fig:geometry}
\end{figure}

We consider two superconductors, denoted 1 and 2, or $L$ (left) and
$R$ (right), connected through a small constriction of diameter
$2a$ at $z=0$, see Fig.~\ref{fig:geometry}. The constriction is assumed to be much
smaller than the superconductor coherence length,
\begin{equation}
2a \ll \xi_{\text{\tiny $\Delta$}} = \frac{v_f}{\pi\Delta}
\,,
\end{equation}
as well as the elastic mean free path, $\ell$, and inelastic mean free
path, $\ell_{\text{\tiny in}}$. We will consider the ballistic limit,
$\ell\gg\xi_{\text{\tiny $\Delta$}}$ in Section~\ref{sec:ballistic}
and the diffusive limit $\ell\ll\xi_{\text{\tiny $\Delta$}}$ in
Section~\ref{sec:diffusive}.  In all cases, we assume that the inelastic
mean free path is much larger than any other length scale in the
problem.

The potential barrier at $z=0$ (the vertical line in
Fig.~\ref{fig:geometry}) is characterized by a transmission
probability, $D(\vp_f)$ for normal-state quasiparticles with Fermi
momentum $\vp_f$ incident on the interface. The reflection probabilty
is given by $R(\vp_f)=1-D(\vp_f)$. The angle-dependence of
$D(\vp_f)$ depends on the microscopic barrier and is not very important
for our purposes; we use a $\delta$-function barrier which gives,
\begin{equation}
D(\vp_f)=\frac{D_0\cos^2\theta}{1-D_0\sin^2\theta},
\end{equation}
where $\theta=\arccos (\hat{\vv}_f\cdot \hat{z})$ is the angle between
quasiparticle trajectory and interface normal, $\vv_f$ is the Fermi
velocity and $D_0$ is the transmission probability for quasiparticles
incident normal to the interface. We develop the theory for heat
conduction for arbitrary transparency of the constriction: $0< D_0
\leq 1$. This theory describes a variety of Josephson weak links,
including the tunnel junction ($D_0\ll 1$) and the pin-hole point contact
($D_0=1$).

For the order parameters of the two superconductors, we restrict
ourselves to the singlet pairing case, but keep the orbital symmetry
arbitrary. Thus, the order parameter is in general assumed to be of
the form $i\sigma_y\Delta(\vp_f)$, where $\sigma_y$ is the second Pauli
matrix in spin space and $\Delta(\vp_f)$ is the momentum dependent gap
amplitude. The results reported below for the heat transport are for
low-$T_c$ superconductors with isotropic order parameters. However,
the formalism is applicable to a broader range of Josephson contacts,
including anisotropic or unconventional spin-singlet pairing states.
We note that current conservation is only guaranteed if the propagators
and self-energies (order parameter, impurity self-energy, etc.) are
calculated self-consistently. In the point-contact geometry we can
neglect the back-action of the current on the self-energies to
lowest order in the small parameter $a/\xi_{\Delta}$.\cite{kul78}

The electrical and thermal resistances of the junction, both $\propto
a^{-2}$, are much larger than the corresponding resistances of the
leads. Thus, the phase change, as well as the temperature change,
occur essentially at the junction. We then take the local phases and
temperatures near the contact to be equal to the reservoir phases and
temperatures. Thus, we write the order parameters as
$\Delta_1(T_1)\,e^{i\phi_1}$ and $\Delta_2(T_2)\,e^{i\phi_2}$ in
superconductor $1$ and $2$, respectively. Here are the local
temperatures near the contact given by $T_1$ and $T_2$, respectively.
We define the phase difference over
the junction $\phi=\phi_2-\phi_1$ and the temperature bias $\Del T=T_2-T_1$.
In the linear response we will write $T_1=T$ and
$T_2=T+\Del T$.

\section{\label{sec:ballistic}Ballistic case}
To study the transport of quasiparticles through superconducting
point contacts as described above, we use the method of nonequilibrium
quasiclassical Green functions.\cite{eli72,lar76,lar77,ser83,ram86} In
this formalism the advanced and retarded Green functions, $\hG^{A,R}$,
describe the local spectrum of excitations of the system, while the
Keldysh Green functions, $\hG^K$, carry information about the
nonequilibrium population of these states. The propagators
$\hG^{A,R,K}(\vp_f,\vR;\epsilon,t)$ are $4\times 4$ matrices in Nambu
(particle-hole and spin) space and obey transport-like equations for
excitations of energy $\epsilon$ moving along classical trajectories
labelled by the Fermi momentum $\vp_f$. The temperature- and phase-biased
contact is, to quasiclassical accuracy, in a stationary state;
thus, we drop the dependence on $t$.

The heat current through the point contact ${\bf
I}_{\epsilon}(\phi,T_1,T_2)$ is a function of the phase difference
$\phi$ and the bath temperatures $T_1$ and $T_2$. In our geometry, the
heat current flows along the junction normal only, ${\bf I}_{\epsilon}
= \hat z \text{I}_{\epsilon}$, and is found by energy integration and
Fermi-surface averaging of the quasiclassical Keldysh Green's
function. Current conservation allows us to write the heat current in
terms of functions locally at the junction on trajectories with
positive and negative projections of the Fermi momentum on the
$\hat\vz$-axis: $\hG^{K}_{+}=\hG^{K}(\vp_f\cdot \hat{z}>0,
\epsilon;\phi,T_1,T_2)$ and $\hG^{K}_{-}=\hG^{K}(\vp_f\cdot \hat{z}<0,
\epsilon;\phi,T_1,T_2)$, respectively. Thus,
\begin{equation}\begin{split}
\text{I}_{\epsilon}(\phi,T_1,T_2) &=
\Area N_f v_f\; \int_{-\infty}^{\infty}
\frac{d\epsilon}{4\pi i}\; \epsilon\\
&\times\left\langle
\text{Tr}\left[ \hG^K_+ -\hG^K_- \right]
\right\rangle,
\end{split}\label{eq:current_def}\end{equation}
where $N_f$ is the normal-state density of states at the Fermi level,
$\Area=\pi a^2$ is the cross-sectional area of the contact, and the
angle brackets denote a Fermi-surface average, including the
projection of the group velocity along the direction normal to
the interface.

A efficient method of computing the Green's functions is provided by
the parametrization in terms of generalized spectral functions
$i\sigma_y\gamma^R$ and distribution functions $x^K$, which are
$2\times 2$ matrices in spin space. In the spin-singlet case, the spin
structure of the spectral functions are all given by $i\sigma_y$,
while the distribution functions are proportional to the unit matrix.
These scalar amplitudes obey Ricatti-type differential
equations.\cite{nag93,sch95,esc99} Each Green's function can be
written in terms of a set of Ricatti amplitudes. The retarded and
advanced Green functions have the forms
\begin{equation}\begin{split}
&\hG^{R,A}[\gamma^{R,A},\tilde\gamma^{R,A}]=\\
&\frac{\mp\pi i}{\zeta^{R,A}}
\left(
\begin{array}{cc}
1-\gamma^{R,A}\tilde\gamma^{R,A} & 2i\sigma_y\gamma^{R,A}\\
-2i\sigma_y\tilde\gamma^{R,A} & -1+\tilde\gamma^{R,A}\gamma^{R,A}
\end{array}
\right),
\end{split}\label{GreenRA}\end{equation}
where $\zeta^{R,A}=1+\gamma^{R,A}\tilde\gamma^{R,A}$, while the the
Keldysh Green function has the form
%
\begin{eqnarray}\label{GreenK}
\hG^K[\gamma^{R},\tilde\gamma^{R},\gamma^A,\tilde\gamma^A,x^K,\tilde x^K]=\qquad\qquad\qquad\qquad
\\
\frac{-2\pi i}{\zeta^R\zeta^A}
\left(
\begin{array}{cc}
x^K+\tilde x^K\gamma^R\tilde\gamma^A & i\sigma_y(x^K\gamma^A-\tilde x^K\gamma^R)\\
-i\sigma_y(x^K\tilde\gamma^R-\tilde x^K\tilde\gamma^A) & \tilde x^K+x^K\tilde\gamma^R\gamma^A
\end{array}
\right)
\nonumber
\,.
\end{eqnarray}
%
The advanced amplitudes are related to the retarded ones through the
symmetry $\gamma^A=-(\tilde\gamma^R)^*$. We will also make use of the
conjugation symmetry\cite{ser83}
\begin{equation}
\tilde q(\vp_f,z,\epsilon)=
q(-\vp_f,z,-\epsilon)^*
\,.\label{eq:sym}
\end{equation}

For each trajectory, the spectral functions and distribution functions
are found by integrating the corresponding Ricatti equations with
initial condition either in the bulk or at the interface, depending on
the stability properties of the differential equation. We follow the
notation in Ref.~\onlinecite{esc00}; quantities denoted by lower case
letters are computed by integrating the Ricatti equations with initial
conditions in the bulk, while quantities denoted with upper case
letters are computed by integrating the Ricatti equations with initial
conditions at the interface.
According to this convention, for each of
the four trajectories in Fig.~\ref{fig:geometry}, we change lower case
letters in Eqs.~\ref{GreenRA}-\ref{GreenK} to upper case letters;
the list of amplitudes defining each Keldysh Green function for the
scattering trajectories are shown in Fig.~\ref{fig:geometry}.

In the bulk region, the amplitudes are given by their equilibrium values
\begin{equation}\label{eq:bulk}\begin{split}
\gamma^R_j(\vp_f,\epsilon) &= -\frac{\Delta_j(\vp_f) e^{i\phi_j}}
{\varepsilon^R+i\sqrt{\Delta_j(\vp_f)^2-(\varepsilon^R)^2}},\\
\tilde{\gamma}^R_j(\vp_f,\epsilon) &= \frac{\Delta_j(\vp_f) e^{-i\phi_j}}
{\varepsilon^R+i\sqrt{\Delta_j(\vp_f)^2-(\varepsilon^R)^2}},\\
x^K_j(\vp_f,\epsilon) &=
(1-|\gamma^R_j(\vp_f,\epsilon)|^2)\tanh {\frac{\epsilon}{2T_j}},\\
\tilde{x}^K_j(\vp_f,\epsilon) &=
-(1-|\tilde{\gamma}^R_j(\vp_f,\epsilon)|^2)\tanh {\frac{\epsilon}{2T_j}},
\end{split}\end{equation}
where the index $j=1,\;2$ refers to quantities in superconductor one
(two). Both $\varepsilon^R$ and the mean field gap $\Delta_j$
formally include renormalization effects (self-energies) from elastic
and inelastic scattering. In the following we consider the ballistic
(clean) case and defer the discussion of impurity scattering to
Sec.~\ref{sec:diffusive}. We also assume that the inelastic scattering
rate, $\Gamma_{\text{\tiny in}}$, is small compared to all relevant energy scales. We
can then write $\varepsilon^R=\epsilon+i\Gamma_{\text{\tiny in}}\rightarrow
\epsilon+i0^+$, and $\Delta_j$ denotes the temperature-dependent
weak-coupling gap amplitude.

The lower case amplitudes are found by integrating the Ricatti
equations from the bulk to the interface, with the initial conditions
in Eqs.~\ref{eq:bulk}. The initial conditions for the upper case Ricatti
amplitudes at the contact are then found from Zaitsev's non-linear
boundary conditions,\cite{zai84} which in terms of Ricatti
amplitudes are reduced to linear boundary conditions.\cite{esc00}
For the distribution functions we use the notation in
Ref.~\onlinecite{lof03c} and obtain,
\begin{equation}\label{X_bc}\begin{split}
X_1^K        &= R_{ee}x_1^K + \bar T_{ee}x_2^K
                + (-\bar T_{eh})\tilde x_2^K,\\
\tilde X_1^K &= R_{hh}\tilde x_1^K + (-\bar T_{he})x_2^K
                + \bar T_{hh}\tilde x_2^K,\\
X_2^K        &= T_{ee}x_1^K + (-T_{eh})\tilde x_1^K
                + \bar R_{ee}x_2^K,\\
\tilde X_2^K &= (-T_{he})x_1^K + T_{hh}\tilde x_1^K
                + \bar R_{hh}\tilde x_2^K.
\end{split}\end{equation}
In terms of the following particle-hole spinor notation,
\begin{equation}
  |\alpha\rangle = \left(
  \begin{array}{c}
    1\\
    -i\sigma_y\alpha
  \end{array}
  \right),\hspace{1cm}
  \langle\alpha| = \left(1,\;\; -i\sigma_y\alpha^* \right),
\end{equation}
the Green's functions at the junction ($z=0$) can be written in a
compact form
\begin{equation}\begin{split}
\hG^K_{1+} &= -\frac{2\pi i}{N_1}
\left[ x_1^K |r_{he}\rangle \langle r_{he}|
+ \tilde X_1^K \hat\tau_1 |\gamma^R_1\rangle
\langle\gamma^R_1|\hat\tau_1 \right],\\
\hG^K_{1-} &= -\frac{2\pi i}{N_2}
\left[ \tilde x_1^K \hat\tau_1|r_{eh}\rangle \langle r_{eh}|\hat\tau_1
+ X_1^K |\tilde\gamma^R_1\rangle \langle \tilde\gamma^R_1| \right],\\
\hG^K_{2-} &= -\frac{2\pi i}{N_3}
\left[ x_2^K |\bar r_{he}\rangle \langle\bar r_{he}|
+ \tilde X_2^K \hat\tau_1|\gamma^R_2\rangle
\langle\gamma^R_2|\hat\tau_1 \right],\\
\hG^K_{2+} &= -\frac{2\pi i}{N_4}
\left[ \tilde x_2^K \hat\tau_1 |\bar r_{eh}\rangle
\langle\bar r_{eh}|\hat\tau_1
+ X_2^K |\tilde\gamma^R_2\rangle \langle \tilde\gamma^R_2|\right],
\end{split}\end{equation}
where $N_k=|\zeta_k|^2$ for $k=1...4$ with
\begin{equation}\begin{aligned}
\zeta_1 &= 1+\gamma^R_1 r_{he}\,, &
\zeta_3 &= 1+\gamma^R_2 \bar r_{he}\,,\\
\zeta_2 &= 1+\tilde\gamma^R_1 r_{eh}\,, &
\zeta_4 &= 1+\tilde\gamma^R_2 \bar r_{eh}\,.
\end{aligned}\end{equation}

The Andreev reflection amplitudes $r_{\alpha\beta}$ are listed in
Table~\ref{ARamplitudes}, and the scattering
probabilities, $T_{\alpha\beta}$ and $R_{\alpha\beta}$, are listed in
Table~\ref{TRprobabilities}. The notation is chosen such that
$T^R_{\alpha\beta}$ ($R^R_{\alpha\beta}$) denotes the probability of
transmission (reflection) of a particle of type $\beta$ to a particle
of type $\alpha$. Quantities with a bar denote transmission from right
to left and reflection on the right side, while those without bar
denote transmission from left to right and reflection on the left
side.

\begin{table}[t]
\caption{The Andreev reflection probabilities. The denominators are
listed in Table~\ref{TRprobabilities}.}
\begin{tabular}{|l|}
\hline
$r_{he} = \tilde \Gamma^R_1 =
[R(1+\tilde\gamma^R_2\gamma^R_2)\tilde\gamma^R_1
+D(1+\tilde\gamma^R_1\gamma^R_2)\tilde\gamma^R_2]/B_1$ \\
$r_{eh} = \Gamma^R_1 = [R(1+\gamma^R_2\tilde\gamma^R_2)\gamma^R_1
+D(1+\gamma^R_1\tilde\gamma^R_2)\gamma^R_2]/B_2$ \\
$\bar r_{he} = \tilde\Gamma^R_2 = [R(1+\gamma^R_1\tilde\gamma^R_1)\tilde\gamma^R_2
+D(1+\tilde\gamma^R_2\gamma^R_1)\tilde\gamma^R_1]/B_3$ \\
$\bar r_{eh} = \Gamma^R_2 = [R(1+\gamma^R_1\tilde\gamma^R_1)\gamma^R_2
+D(1+\gamma^R_2\tilde\gamma^R_1)\gamma^R_1]/B_4$\\
\hline
\end{tabular}\label{ARamplitudes}
\end{table}
\begin{table*}[t]
\caption{Scattering probabilities for $x^K$ distribution functions in
the stationary SIS junction setup.}
\begin{tabular}{|l|l|l|l|}
\hline
$\bar T_{hh} = D|1+\tilde\gamma^R_1\gamma^R_2|^2/|B_1|^2$ &
$\bar T_{he} = RD|\tilde\gamma^R_1-\tilde\gamma^R_2|^2/|B_1|^2$ &
$R_{hh} = R|1+\tilde\gamma^R_2\gamma^R_2|^2/|B_1|^2$ &
$B_1 = 1+R\gamma^R_2\tilde\gamma^R_2+D\tilde\gamma^R_1\gamma^R_2$\\
\hline
$\bar T_{ee} = D|1+\gamma^R_1\tilde\gamma^R_2|^2/|B_2|^2$ &
$\bar T_{eh} = RD|\gamma^R_1-\gamma^R_2|^2/|B_2|^2$ &
$R_{ee} = R|1+\gamma^R_2\tilde\gamma^R_2|^2/|B_2|^2$ &
$B_2 = 1+R\gamma^R_2\tilde\gamma^R_2+D\gamma^R_1\tilde\gamma^R_2$ \\
\hline
$T_{hh} = D|1+\tilde\gamma^R_2\gamma^R_1|^2/|B_3|^2$ &
$T_{he} = RD|\tilde\gamma^R_2-\tilde\gamma^R_1|^2/|B_3|^2$ &
$\bar R_{hh} = R|1+\gamma^R_1\tilde\gamma^R_1|^2/|B_3|^2$ &
$B_3 = 1+R\gamma^R_1\tilde\gamma^R_1+D\tilde\gamma^R_2\gamma^R_1$ \\
\hline
$T_{ee} = D|1+\gamma^R_2\tilde\gamma^R_1|^2/|B_4|^2$ &
$T_{eh} = RD|\gamma^R_2-\gamma^R_1|^2/|B_4|^2$ &
$\bar R_{ee} = R|1+\gamma^R_1\tilde\gamma^R_1|^2/|B_4|^2$ &
$B_4 = 1+R\gamma^R_1\tilde\gamma^R_1+D\gamma^R_2\tilde\gamma^R_1$ \\
\hline
\end{tabular}\label{TRprobabilities}
\end{table*}

We use the above interface Green functions to compute the heat current
via Eq.~\ref{eq:current_def}. There are four contributions, one for
each {\it incoming} distribution function in Eq.~\ref{eq:bulk}:
$i=1-4$ for $x^K_1,\tilde x^K_1,x^K_2,\tilde x^K_2$, corresponding to
electron-like and hole-like quasiparticles injected from the left
($i=1-2$) and right ($i=3-4$) reservoirs,
\begin{equation}
\text{I}_{\epsilon} = \int_{-\infty}^{\infty}d\epsilon\left\langle
\sum_{i=1}^{4} j_{\epsilon}^i({\bf p}_f,z,\epsilon;\phi)\right\rangle
\,.
\end{equation}
We can express the spectral current in terms of
transmission probabilities, $T^R_{\alpha\beta}$, for the distribution
functions. The most direct way of
achieving this is to make use of the unitarity of the scattering
matrix for the junction, which leads to the relation
\begin{equation}\begin{split}
&\text{Tr}\left[ \hG^K_+({\bf p}_f,z=0^-,\epsilon)
-\hG^K_-({\bf p}_f,z=0^-,\epsilon) \right]\\
&=\text{Tr}\left[ \hG^K_+({\bf p}_f,z=0^+,\epsilon)
-\hG^K_-({\bf p}_f,z=0^+,\epsilon) \right].
\end{split}\end{equation}
Thus, we compute $j_{\epsilon}^1$ and $j_{\epsilon}^2$ at $z=0^+$,
while $j_{\epsilon}^3$ and $j_{\epsilon}^4$ are computed at
$z=0^-$. The heat current spectral densities can then be written
as

\begin{widetext}
\begin{equation}\begin{split}
j_{\epsilon}^1({\bf p}_f,z=0^+,\epsilon;\phi,T_1,T_2) &=
-\Area N_f v_f \epsilon\, x^K_1 \left[
T_{ee} (1-|\tilde\gamma^R_2|^2) N_4^{-1}
+T_{he} (1-|\gamma^R_2|^2) N_3^{-1} \right],\\
j_{\epsilon}^2({\bf p}_f,z=0^+,\epsilon;\phi,T_1,T_2) &=
-\Area N_f v_f \epsilon\, \tilde x^K_1 \left[
-T_{eh} (1-|\tilde\gamma^R_2|^2) N_4^{-1}
-T_{hh} (1-|\gamma^R_2|^2) N_3^{-1} \right],\\
j_{\epsilon}^3({\bf p}_f,z=0^-,\epsilon;\phi,T_1,T_2) &=
-\Area N_f v_f \epsilon\, x^K_2 \left[
-\bar T_{he} (1-|\gamma^R_1|^2) N_1^{-1}
-\bar T_{ee} (1-|\tilde\gamma^R_1|^2) N_2^{-1} \right],\\
j_{\epsilon}^4({\bf p}_f,z=0^-,\epsilon;\phi,T_1,T_2) &=
-\Area N_f v_f \epsilon\, \tilde x^K_2 \left[
\bar T_{hh} (1-|\gamma^R_1|^2) N_1^{-1}
+\bar T_{eh} (1-|\tilde\gamma^R_1|^2) N_2^{-1} \right].
\end{split}\end{equation}
\end{widetext}

Note that each spectral current density vanishes in the subgap region, i.e. for
$|\epsilon|<\max(\Delta_1,\Delta_2)$ the factor
$1-|\gamma^R|^2=0$. The heat current is only carried by continuum
energy quasiparticles, and thus, we only consider
$|\epsilon|>\max(\Delta_1,\Delta_2)$ from here on. Using the symmetry
in Eq.~\ref{eq:sym} we obtain
\begin{equation}\label{eh_sym}\begin{split}
j_{\epsilon}^2({\bf p}_f,z=0^+,\epsilon) &=
j_{\epsilon}^1(-{\bf p}_f,z=0^+,-\epsilon),\\
j_{\epsilon}^4({\bf p}_f,z=0^-,\epsilon) &=
j_{\epsilon}^3(-{\bf p}_f,z=0^-,-\epsilon),
\end{split}\end{equation}
which implies that hole-like quasiparticles carry the same amount of
heat as the electron-like quasiparticles (after energy integration and
Fermi-surface averaging).

We introduce transmission coefficients (script $\D$'s) by
combining the scattering probabilities for $x^K$ distribution
functions (big $T$'s) with the spectral renormalization factors [the factors,
$(1-|\gamma^R|^2)$, and the denominators $N^{-1}_i$]. We
then obtain,
\begin{equation}\begin{split}
\hspace*{-3mm}j_{\epsilon}^1 &=
-\Area N_f v_f \epsilon\, \tanh\frac{\epsilon}{2T_1}
\left[\D_{ee}({\bf p}_f,\epsilon)
+\D_{he}({\bf p}_f,\epsilon)\right],\\
\hspace*{-3mm}j_{\epsilon}^3 &=
+\Area N_f v_f \epsilon\, \tanh\frac{\epsilon}{2T_2}
\left[\bar\D_{ee}({\bf p}_f,\epsilon)
+\bar\D_{he}({\bf p}_f,\epsilon)\right].
\end{split}\end{equation}
The transmission coefficient for electron-like quasiparticles
remaining electron-like, $\D_{ee}$, has the form
\begin{equation}\begin{split}
\D_{ee}({\bf p}_f,\epsilon) &=
(1-|\gamma^R_1|^2)T_{ee}(1-|\tilde\gamma^R_2|^2)N_4^{-1}\\
&=\frac{D(1-|\gamma^R_1|^2)(1-|\tilde\gamma^R_2|^2)|1+\gamma^R_2\tilde\gamma^R_1|^2}
{|Z|^2},
\end{split}\end{equation}
while the transmission coefficient for electron-like quasiparticles
with branch conversion to hole-like quasiparticles, $\D_{he}$, is
\begin{equation}
\hspace*{-5mm}\D_{he}({\bf p}_f,\epsilon) =
\frac{RD(1-|\gamma^R_1|^2)(1-|\gamma^R_2|^2)|\tilde\gamma^R_2-\tilde\gamma^R_1|^2}
{|Z|^2}.
\end{equation}
The corresponding coefficients for transmission from right to left are,
\begin{equation}\begin{split}
\hspace*{-5mm}\bar\D_{ee}({\bf p}_f,\epsilon) &=
\frac{D(1-|\gamma^R_2|^2)(1-|\tilde\gamma^R_1|^2)|1+\gamma^R_1\tilde\gamma^R_2|^2}
{|Z|^2},\\
\hspace*{-5mm}\bar\D_{he}({\bf p}_f,\epsilon) &=
\frac{RD(1-|\gamma^R_2|^2)(1-|\gamma^R_1|^2)|\tilde\gamma^R_1-\tilde\gamma^R_2|^2}
{|Z|^2}.
\end{split}\end{equation}
The common denominator is given by
\begin{equation}\begin{split}
Z =& 1+R(\gamma^R_1\tilde\gamma^R_1+\gamma^R_2\tilde\gamma^R_2)\\
&+D(\gamma^R_1\tilde\gamma^R_2+\tilde\gamma^R_1\gamma^R_2)
+\gamma^R_1\tilde\gamma^R_1\gamma^R_2\tilde\gamma^R_2.
\end{split}\end{equation}
We immediately see that
\begin{equation}
\bar\D_{he}(\vp_f,\epsilon)=\D_{he}(\vp_f,\epsilon),
\end{equation}
while the symmetry given by Eq.~\ref{eq:sym} leads to
\begin{equation}
\bar\D_{ee}(\vp_f,\epsilon)=\D_{ee}(-\vp_f,-\epsilon).
\end{equation}
With the aid of these relations we can express the total heat current
in a more compact form,
\begin{equation}\label{HeatCurrent}
\text{I}_{\epsilon}(\phi,T_1,T_2) =
2\int_{\max(\Delta)}^{\infty}d\epsilon\;
j_{\epsilon}(z=0,\epsilon;\phi,T_1,T_2),
\end{equation}
where the factor $2$ reflects the symmetry of the spectral current
under $-\epsilon\to\epsilon$.
The heat-current spectral density is expressed in terms
of the Fermi-surface averaged transmission coefficient,
\begin{equation}\label{HeatCurrentDens}\begin{split}
j_{\epsilon}(z=0,\epsilon;\phi,T_1,T_2) &= -2\Area N_f v_f \epsilon\;
\Big\langle\D({\bf p}_f,\epsilon,\phi;T_1,T_2)\Big\rangle\\
&\quad\left[\tanh\frac{\epsilon}{2T_1}-\tanh\frac{\epsilon}{2T_2}\right]
\,,
\end{split}\end{equation}
\begin{equation}
\D({\bf p}_f,\epsilon,\phi) =
\D_{ee}({\bf p}_f,\epsilon,\phi)+\D_{he}({\bf p}_f,\epsilon,\phi)
\,,
\end{equation}
where the factor two is due to the electron-hole symmetry
[c.f. Eq.~\ref{eh_sym}]. Equation \ref{HeatCurrent} satisfies the symmetries
\begin{equation}\begin{split}
\text{I}_{\epsilon}(-\phi,T_1,T_2) &= \text{I}_{\epsilon}(\phi,T_1,T_2),\\
\text{I}_{\epsilon}(\phi,T_2,T_1) &= -\text{I}_{\epsilon}(\phi,T_1,T_2).
\end{split}\end{equation}
and Eqs.~\ref{HeatCurrent}-\ref{HeatCurrentDens} are the
results for non-linear heat-current response to a temperature and
phase bias. These results hold for singlet superconductors with any orbital
symmetry, in particular for either $s$-wave or $d$-wave symmetries.

\subsection{S-wave symmetry: $\Delta({\bf p}_f)=\Delta$}

In the remaining sections we focus on the low-$T_c$ superconductors, with
a momentum independent ($s$-wave) order parameter. However, the order
parameters for the two superconductors have different phases, $\phi_j$, and
may also have different order parameter amplitudes,
e.g. as a result of a finite temperature bias, $\Delta_j=\Delta(T_j)$.
The retarded Ricatti amplitudes in the bulk
are then momentum independent. Moreover, since the s-wave order parameter is
constant in space for the point contact, the lower case Ricatti amplitudes
are given by their bulk values along the whole
trajectory. We then obtain analytic expressions for the effective
transmission coefficients
\begin{equation}\label{eq:D-swave}\begin{split}
\hspace*{-3mm}\D_{ee}({\bf p}_f,\epsilon) &=
\frac{2D\xi_1\xi_2(\epsilon^2+\xi_1\xi_2-\Delta_1\Delta_2\cos\phi)}
{\left[D\epsilon^2+(1+R)\xi_1\xi_2-D\Delta_1\Delta_2\cos\phi\right]^2},\\
\hspace*{-3mm}\D_{he}({\bf p}_f,\epsilon) &=
\frac{2RD\xi_1\xi_2(\epsilon^2-\xi_1\xi_2-\Delta_1\Delta_2\cos\phi}
{\left[D\epsilon^2+(1+R)\xi_1\xi_2-D\Delta_1\Delta_2\cos\phi\right]^2}.
\end{split}\end{equation}
where
$\xi_j\equiv\text{sgn}(\epsilon)\sqrt{\epsilon^2-\Delta_j^2}$.

\subsection{Linear response}

In linear response we write $T_1=T$ and $T_2=T+\delta T$. Then to
lowest order in $\delta T$ we obtain,
\begin{equation}
I_{\epsilon}(\phi,T) = -\kappa(\phi,T)\,\delta T,\label{eq:heat-linear}
\end{equation}
where the heat conductance has the form
\begin{equation}\label{eq:conductance}
\hspace*{-4mm}\kappa(\phi,T) = 4\Area \int_{\Delta}^{\infty}d\epsilon
\cN(\epsilon)
[\epsilon v_g(\epsilon)]
\left\langle \D(\epsilon,\phi,T)\right\rangle\,
\left(\frac{\partial f}{\partial T}\right),
\end{equation}
with the transmission coefficient,
\begin{equation}\label{eq:D-linear}
\hspace*{-5mm}\D({\bf p}_f,\epsilon,\phi,T)=\D_{ee}({\bf p}_f,\epsilon,\phi,T)
                               + \D_{he}({\bf p}_f,\epsilon,\phi,T)
\,,
\end{equation}
\begin{equation}\begin{split}
\D_{ee}({\bf p}_f,\epsilon,\phi,T) &=
D\frac{(\epsilon^2-\Delta^2)(\epsilon^2-\Delta^2\cos^2\frac{\phi}{2})}
{\left[\epsilon^2-\Delta^2(1-D\sin^2\frac{\phi}{2})\right]^2},\\
\D_{he}({\bf p}_f,\epsilon,\phi,T) &=
RD\frac{(\epsilon^2-\Delta^2)\,\Delta^2\sin^2\frac{\phi}{2}}
{\left[\epsilon^2-\Delta^2(1-D\sin^2\frac{\phi}{2})\right]^2}.
\end{split}
\end{equation}

Equation~\ref{eq:conductance} is an intuitive form of the heat
conductance that resembles the bulk
thermal conductivity. The conductance is expressed in terms of the bulk
quasiparticle density of states,
$\cN(\epsilon)=N_f|\epsilon|/\sqrt{\epsilon^2-\Delta^2}$, and the energy current
carried by these quasiparticles, $[\epsilon\,v_g(\epsilon)]$, where
$v_g(\epsilon)=v_f\sqrt{\epsilon^2-\Delta^2}/|\epsilon|$ is the group velocity
of a bulk excitation. Note that the product, $\cN v_g = N_fv_f$, is energy
independent. The backscattering at the junction (resulting from the Sharvin
resistance and the transparency $D\leq 1$), which limits the heat conductance,
corresponds to the elastic mean free path due to impurities in the bulk.

The linear response results in Eqs.~\ref{eq:heat-linear}-\ref{eq:D-linear} for the
heat conductance were reported in Ref.~\onlinecite{zha03c}, for a
$\vp_f$-independent normal-state transmission probability, $D$.
In Sec.~\ref{sec:linear} we extend the analysis of the linear response limit
and examine the non-linear thermal response, Eqs.~\ref{HeatCurrent}-\ref{HeatCurrentDens},
in Sec.~\ref{sec:nonlinear}.

\section{Thermal Conductance}\label{sec:linear}

\begin{figure}[t]
\includegraphics[width=8cm]{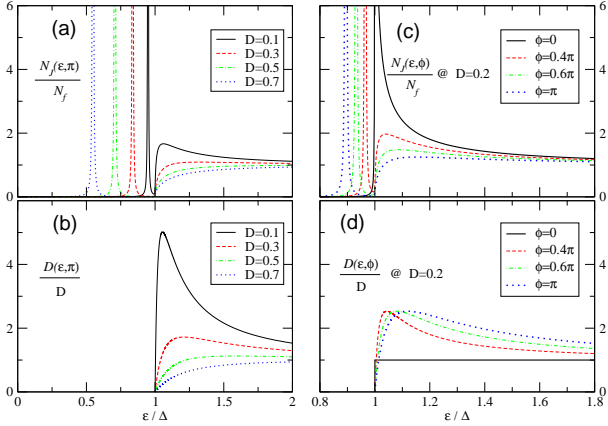}
\caption{The LDOS at the junction,
$N_J(\epsilon,\phi)$, and the transmission coefficient,
$\D(\epsilon,\phi)$, at fixed momentum $\vp_f$ with
$D\equiv D(\vp_f)$. (a)-(b) Different transparencies at phase difference
$\phi=\pi$. (c)-(d) Different phase differences at transparency
$D=0.2$. $N_J$ is suppressed by the formation of the ABS, as
$\phi$ is tuned from $0$ to $\pi$. For high transparency $\D$ is
suppressed, but for low transparency $\D$ contains a resonance
at $\epsilon_r\simeq\Delta+\frac{1}{2}\Delta
D\sin^2\frac{\phi}{2}$, at which $\D(\epsilon_r,\phi)=1/2$. In all
cases, $T=0.72T_c$, and for clarity we added a small
width, $\sim 10^{-3}T_c$ to the bound-states.}
\label{fig:LDOS_Dtr_lin}
\end{figure}

The phase modulation of the heat conductance, $\kappa(\phi,T)$, is
determined by the transmission coefficient, $\D(\epsilon,\phi,T)$.
The energy dependence of the transmission coefficient reflects features
of the phase-dependent local density of states (LDOS) of the junction,
$\cN_J(\epsilon,\phi,T)$. The LDOS at the contact is defined as
$
\cN_J(\vp_f,\epsilon,\phi,T)=-\frac{N_f}{\pi}\text{Im}\hG^R_{11}(\vp_f,z=0^-,\epsilon,\phi,T)
$,
where $\hG^R_{11}$ is the diagonal (11) component of the retarded Green's function,
which has the following form at the junction,
\begin{equation}
\frac{\hG^R_{11}(\vp_f,z=0^-,\epsilon)}{-i\pi} =
\frac{\epsilon\sqrt{\Delta^2-\epsilon^2}
-\frac{i}{2}\text{sgn}(\vp_f\cdot\hat\vz)\Delta^2 D\sin\phi}
{\epsilon^2-\Delta^2(1-D\sin^2\frac{\phi}{2})}
\,.
\end{equation}
The resulting LDOS has both a bound-state ($|\epsilon|<\Delta$) and
continuum ($|\epsilon|>\Delta$) spectrum given by
\begin{widetext}
\begin{equation}\begin{aligned}
\frac{\cN_J(\vp_f,\epsilon,\phi,T)}{\cN_f} &= \Theta(|\epsilon|-\Delta)
\frac{|\epsilon|\sqrt{\epsilon^2-\Delta^2}}
{\epsilon^2-\Delta^2(1-D\sin^2(\phi/2))}\\
&\quad+\pi\delta(\epsilon-\epsilon_{B})
\frac{|\epsilon|\sqrt{\Delta^2-\epsilon^2}
-\frac{1}{2}\text{sgn}(\vp_f\cdot\hat\vz)\Delta^2D\sin\phi}
{2|\epsilon|} +\pi\delta(\epsilon+\epsilon_{B})
\frac{|\epsilon|\sqrt{\Delta^2-\epsilon^2}
+\frac{1}{2}\text{sgn}(\vp_f\cdot\hat\vz)\Delta^2D\sin\phi}
{2|\epsilon|},
\end{aligned}
\end{equation}
\end{widetext}
The spectrum is shown in Fig.~\ref{fig:LDOS_Dtr_lin} as a function of the phase
bias (\ref{fig:LDOS_Dtr_lin}c) and as a function of barrier transparency for
$\phi=\pi$ (\ref{fig:LDOS_Dtr_lin}a). Of particular importance is the formation
of a pair of Andreev bound states (ABS) with phase dispersion
\begin{equation}
\epsilon_B = \pm\Delta\sqrt{1-D\sin^2\frac{\phi}{2}}
\,,
\end{equation}
and the impact of the ABS on the continuum spectrum.
In addition, we note the following characteristics of the LDOS:
\begin{enumerate}
\item In the absence of phase bias, $\phi=0$, the LDOS reduces to the
      BCS density of states in the bulk.
\item Under a phase bias, $\phi\neq 0$, Andreev bound states are formed,
      with spectral weight drawn from the continuum near $\epsilon=\Delta$.
\item The ABS are weakly bound and close to the continuum edge for
      low transparency. For high transparency these states are more strongly bound, and may lie
      well below the gap.
\item For $\phi=\pi$ the ABS is the closest to the Fermi level (at
      $\epsilon_B=\Delta\sqrt{R}$), and consequently the spectral weight of
      continuum excitations is reduced the most at this phase bias.
\end{enumerate}
These characteristics of the LDOS have direct consequences for the properties of
the transmission coefficient, $\D$, and thus for the heat conductance.
For $\phi=0$ the transmission coefficient becomes independent of energy
and reduces to the normal-state transmission probability:
$\D(\epsilon,0)=D$. Thus, the thermal conductance of the point contact
at $\phi=0$,
\begin{equation}\label{kappa_phi=0}
\kappa(\phi=0)=\frac{N_f v_f\,\langle D\rangle\,\Area}{T^2}
\int_{\Delta}^{\infty} d\epsilon\epsilon^2\mathrm{sech}^2\frac{\epsilon}{2T}
\,,
\end{equation}
is reduced compared to the normal-state conductance (Eq.~\ref{kappa_phi=0} with $\Delta=0$),
$\kappa_{\text{N}}=\frac{2\pi^2}{3}N_f v_f\,\langle D\rangle\,\Area\,T$,
by opening the superconducting gap. Note that the temperature dependence of the ratio,
$\kappa(\phi=0)/\kappa_{\text{N}}(T_c)$,
is equivalent to that of the normalized bulk thermal
conductivity of an s-wave BCS superconductor.
\begin{figure}[t]
\includegraphics[height=2.4in]{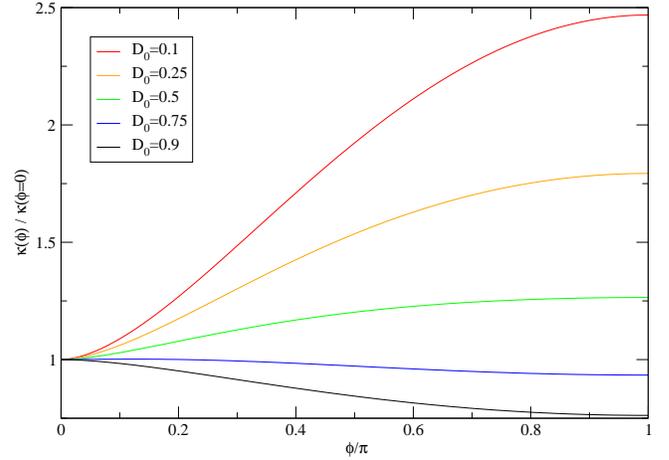}
\caption{The thermal conductance as a function of $\phi$ and barrier
transparency $D$.  The thermal conductance is normalized for each $D$
by its value at $\phi=0$.}\label{fig:kappa-phase}
\end{figure}
For $\phi\neq 0$, the transmission coefficient is strongly
dependent on energy and phase, which is the source of the phase-modulation of the
thermal conductance shown in Fig.~\ref{fig:kappa-phase}. For high
transparency ($D\gtrsim 0.7$), the suppression of the continuum density of states
associated with the formation of the ABS well below the gap is
reflected in a suppressed transmission coefficient and a reduced
heat conductance compared to $\phi=0$. Thus, for high transparency as $\phi$ is tuned
from $0$ to $\pi$, the conductance is suppressed as shown in
Fig.~\ref{fig:kappa-phase}. In the limit $D\rightarrow 1$ only direct
transmission is possible,
$\D_{ee}(\epsilon,\phi)=(\epsilon^2-\Delta^2)/(\epsilon^2-\Delta^2\cos^2(\frac{\phi}{2}))$;
branch conversion processes vanish, $\D_{he}=0$, and we recover the
heat conductance of a pinhole junction
\begin{equation}
\kappa(\phi)=\frac{N_f v_f \Area}{4 T^2} \int_{\Delta}^{\infty} d\epsilon
\frac{\epsilon^2(\epsilon^2-\Delta^2)}{\epsilon^2-\Delta^2\cos^2\frac{\phi}{2}}
\mathrm{sech}^2\frac{\epsilon}{2T}\,,
\end{equation}
obtained by Kulik and Omelyanchuk.\cite{kul92}

In the case of low transparency, $D\ll 1$, and $\phi\neq 0$,
the Andreev bound states lie just below the
gap edge. The transmission coefficient,
$\D(\epsilon,\phi)$, exhibits a resonance just {\it above} the gap
edge, as shown in Fig.~\ref{fig:LDOS_Dtr_lin}. This resonance is
a consequence of the shallow ABS. We see this
fact clearly in Fig.~\ref{fig:LDOS_Dtr_lin}(c)-(d): the continuum LDOS
is reduced as $\phi$ is changed from $0$ to $\pi$, while the
resonance in the transmission coefficient is enhanced (the area under
the curve is enhanced). The resonance energy is obtained
by writing $\epsilon_r=\Delta+\delta\epsilon_r$, where
$\delta\epsilon_r\ll\Delta$, in the expression for the transmission
coefficient, Eq.~\ref{eq:D-linear}. From the extremum condition,
$\partial\D/\partial\,\delta\epsilon_r=0$, we obtain
\begin{equation}
\epsilon_r \approx \Delta\left(1+\frac{1}{2} D \sin^2\frac{\phi}{2}\right).
\end{equation}
At the resonance energy the transmission coefficient is independent of
transparency and takes the value
\begin{equation}
\D(\epsilon_r,\phi)=\frac{1}{2}.
\end{equation}
The resonant transmission of quasiparticles with energies
$\epsilon\approx\epsilon_{\text{r}}$
leads to an increase in the thermal conductance as the phase bias is
tuned from $0$ to $\pi$, as shown in Fig.~\ref{fig:kappa-phase}.
Observation of the phase modulation may be accomplished with a point-contact
Josephson junction in a SQUID geometry. In this case the phase is tuned by the flux, $\Phi$,
threading the SQUID, $\phi=2\pi(\Phi/\Phi_0)$, where $\Phi_0=hc/2e$ is the superconducting
flux quantum.

\begin{figure}
\includegraphics[height=2.4in]{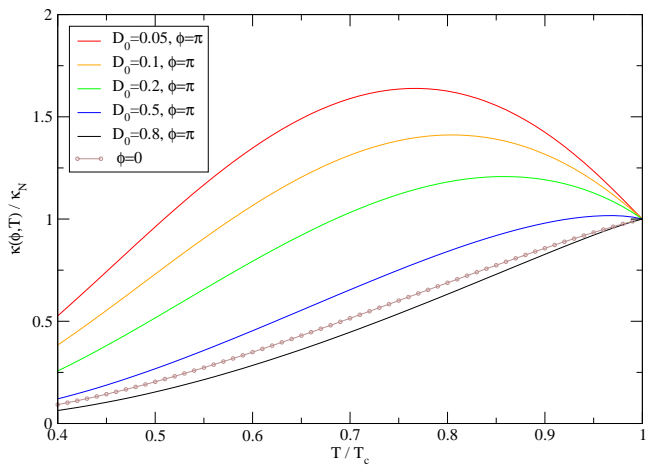}
\caption{The temperature dependence of phase-modulation of the thermal
conductance for $\phi=\pi$, normalized by the conductance at $T_c$,
$\kappa_N=2\pi^2\Area N_f v_f T_c \left<D\right>/3$. Shown for
comparison is the normalized conductance for $\phi=0$, which is
independent of transparency.}
\label{fig:kappa-T}
\end{figure}

Another important consequence of the resonance is the {\it increase} of the thermal
conductance for $\phi=\pi$ compared to the normal-state conductance at
$T_c$, as the temperature is lowered below $T_c$, see Fig.~\ref{fig:kappa-T}.
This peak effect is large and persists
over a broad temperature range, $0.5T_c \lesssim T \le T_c$, for moderate to low
transparencies, $D\sim 0.2$ to $D\ll 1$.

For very low temperatures, $T\ll\Delta$, the sharpening of the distribution
functions at the Fermi level, combined with the gap in the continuum
spectrum leads to an exponentially small thermal conductance.
The reduction of the low-temperature heat
conductance for normal-superconducting (NS) proximity structures
is well known from Andreev's work on heat reflection at NS interfaces.\cite{and64}
However, the {\it increase} in the thermal conductance compared to the normal state
in the intermediate to high temperature range
is not related to Andreev reflection at high-transparency interfaces,
but is a consequence resonant transmission of heat carrying quasiparticles, and
is characteristic of intermediate- to low-transparency Josephson point-contact
junctions.

\subsection{The tunnelling limit: $D\to 0$}\label{sec:tunnelconductance}

\begin{figure}
\includegraphics[height=2.4in]{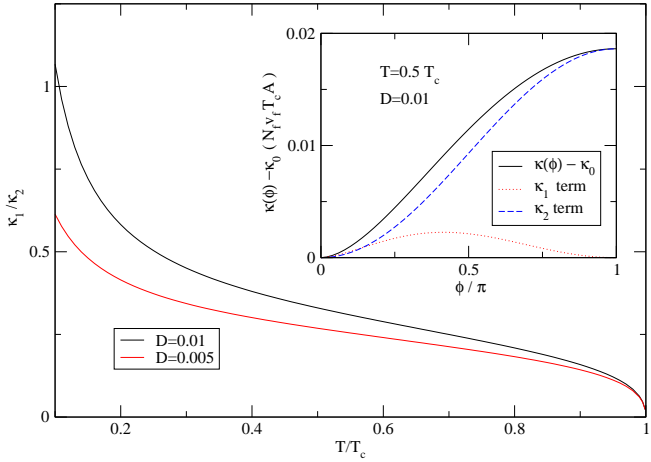}
\caption{The ratio $\kappa_1/\kappa_2$ as a function of temperature.
The inset shows the phase dependence of the thermal conductance,
the contributions from terms proportional to
$\kappa_1$ and $\kappa_2$ are plotted separately.}
\label{fig:kappa}
\end{figure}

Here we examine in more detail the phase dependence of the thermal conductance
in Eq.~\ref{eq:conductance} in the tunnelling limit, $D\ll 1$. We neglect the
momentum dependence and set $D(\vp_f)=D$. For $D\ll 1$, when we expand the
transmission coefficient, $\D(\epsilon,\phi)$, to leading order in $D$
there is prefactor proportional to $D$, and a non-perturbative
dependence on $D$ that is responsible for the resonance peak at
$\epsilon_{\text{r}}$. The resonance leads to a non-analytic contribution to the
conductance proportional $D\ln D$ that dominates in the limit $D\to 0$.
Consequently up to order $D$ the thermal conductance $\kappa(\phi)$
contains a term proportional to $\sin^2{\phi\over 2}$, as well as a nonanalytic term
proportional to $\sin^2{\phi\over 2}\ln(\sin^2{\phi\over 2})$,
\begin{equation}
\kappa(\phi)=\kappa_0-\kappa_1\sin^2\frac{\phi}{2}
\ln(\sin^2\frac{\phi}{2})+\kappa_2 \sin^2\frac{\phi}{2} \,,
\label{eq:conductance_small_D}
\end{equation}
where $\kappa_0$ is given by Eq.~\ref{kappa_phi=0} and
\[
\kappa_1={N_f v_f D\Area \Delta^3 \over 4T^2}
\mathrm{sech}^2{\Delta\over 2T},
\]
\[
\kappa_2={N_f v_f D\Area \Delta^3 \over 4T^2} \left[{4T\over
\Delta}(1-\tanh{\Delta\over 2T})+c\right]-(1+\ln D)\kappa_1,
\]
\[
c=2\int_0^{\infty}dx x\ln x \left[ { {\Delta\over
T}\;\mathrm{sinh}{\sqrt{x^2+1}\over 2T/\Delta} \over (x^2+1)
\mathrm{cosh}^3{\sqrt{x^2+1}\over 2T/\Delta}} +
{\mathrm{sech}^2{\sqrt{x^2+1}\over 2T/\Delta} \over
(x^2+1)^{3/2}}\right].
\]
The phase modulation of the thermal conductance is shown in the inset
of Fig.~\ref{fig:kappa}, the term $\kappa_2\sin^2{\phi\over 2}$ and
the term $-\kappa_1\sin^2{\phi\over 2}\ln(\sin^2{\phi\over 2})$ are
plotted separately for $T=0.5T_c$ and $D=0.01$. The relative
importance of the two terms is shown for
$D=0.01$ and $D=0.005$ as a function of temperature. The ratio
$\kappa_1/\kappa_2$ increases as temperature is lowered.

We compare Eq.~\ref{eq:conductance_small_D} with the result obtained
with the tunnelling Hamiltonian (tH) method
(see Refs.~\onlinecite{gut97,gut98} and Appendix). According to the tH
calculation, the heat transport through a tunnel junctions is given by
Eq.~\ref{int} in the Appendix. In the linear response limit,
$\delta T\rightarrow 0$, $\Delta_1=\Delta_2$, and Eq.~\ref{int} reduces to
\begin{equation}
\kappa^{\mathrm{tH}}(\phi)=8\pi N_f^2 |{\bf T}|^2
\int^{+\infty}_{\Delta} d\epsilon \;\epsilon
{\partial f(\epsilon) \over \partial T}
{\epsilon^2-\Delta^2 \cos\phi \over
\epsilon^2-\Delta^2 } .\label{tH}
\end{equation}
The integral in Eq.~\ref{tH} is divergent due to the
singularity at $\epsilon=\Delta$. The divergence is unphysical and
indicative of the failure of low-order perturbation theory within the tH
method for the linear response.
What is missing in Eq.~\ref{tH} is the correction to the spectrum - which
enters the denominator -
that results from the small, but finite, barrier transparency. This
non-perturbative correction is fully taken into account in the
quasiclassical Green function method. The singularity of the DOS at
$\epsilon=\Delta$ is removed by the formation of the ABS, and the
final result, Eqs.~\ref{eq:conductance}-\ref{eq:D-linear},
contain no divergence in the linear response limit.

In earlier treatments the divergence entering the tH result was
regulated by introducing an ad-hoc cutoff,\cite{gut97,gut98} for
example by requiring that the two gaps have different magnitudes,
$\Delta_1\neq \Delta_2$. Under these circumstances, Eq.~\ref{int}
predicts the thermal conductance has the form
$\kappa(\phi)=\kappa'_0-\kappa'_2\cos\phi$, with both $\kappa'_0$ and
$\kappa'_2$ proportional to $|{\bf T}|^2\propto D$. However, as shown in
Eq.~\ref{eq:conductance_small_D} the thermal conductance of a tunnel
junction contains non-perturbative corrections to the cosine
dependence, i.e. the leading order correction is
$-\kappa_1\sin^2{\phi\over 2}\ln(\sin^2{\phi\over 2})$. Also the
magnitude of $\kappa_2$ contains
a term proportional to $D\ln D$.

\section{Non-linear response}\label{sec:nonlinear}

\begin{figure}
\includegraphics[width=8cm]{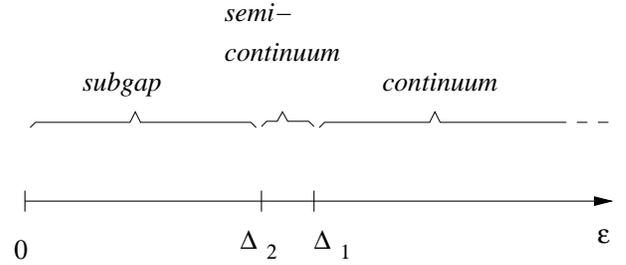}
\caption{Three regions in energy: \textsl{sub-gap} energies below the gap of
both superconducting leads, \textsl{semi-continuum} energies between the two
gap energies, and the true \textsl{continuum}. Only the true continuum states
transport heat.}
\label{fig:spectrum}
\end{figure}

\begin{figure}[t]
\includegraphics[width=8cm]{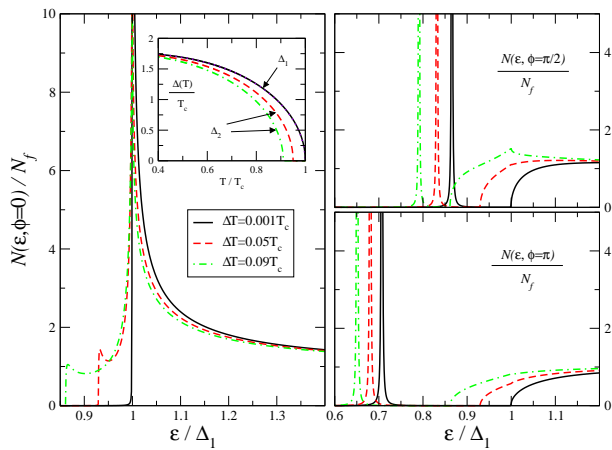}
\caption{Local density of states at the junction at $z=0^-$ in the
non-linear response, for transparency $D=0.5$ and temperature
$T=0.72T_c$.}\label{fig:ldos_nonlinear}
\end{figure}

When we consider the non-linear response one important effect is
that the order parameters on the two sides have
different magnitudes,
\begin{equation}\begin{split}
\Delta_1 &= \Delta(T),\\
\Delta_2 &= \Delta(T+\Del T)<\Delta_1.
\end{split}\end{equation}
The density of states is modified accordingly. There are three
important regions of the local excitation spectrum (c.f. Fig.~\ref{fig:spectrum}):
\begin{enumerate}
\item $|\epsilon|<\Delta_2$, sub-gap spectrum with bound states,
\item $\Delta_2<|\epsilon|<\Delta_1$, semi-continuum spectrum,
\item $\Delta_1<|\epsilon|$, true continuum spectrum.
\end{enumerate}
Only the states in the true continuum carry heat. In
Fig.~\ref{fig:ldos_nonlinear} we plot the local density of states at the
junction for three different phase differences. The difference between the
gap magnitudes becomes large as $\Del T$ is increased, the bound states
move closer to the semi-continuum edge. At the same time, the true continuum
edge is further separated from the ABS. Consequently, the true continuum is
less affected by the bound states, and therefore the phase bias.

\begin{figure}[t]
\includegraphics[width=8cm]{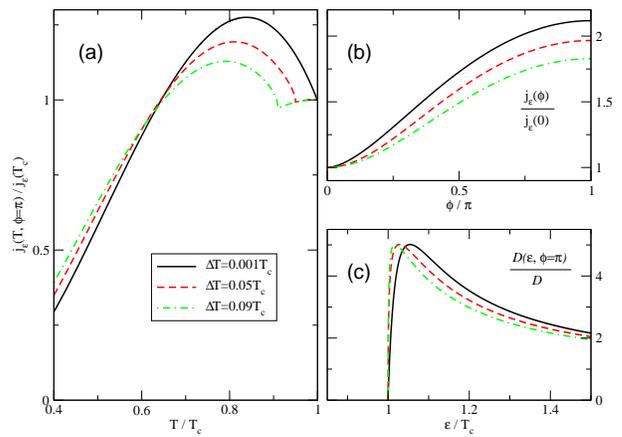}
\caption{(a) Temperature dependence and (b) phase dependence of the
heat current in the non-linear response. (c) Transmission coefficient
in the non-linear response at phase bias $\phi=\pi$. In (b)-(c) the
temperature is $T=0.72T_c$, and in all cases the transparency of the
junction is $D=0.1$.}\label{fig:jE_nonlinear}
\end{figure}

The Andreev bound states also have an impact on the heat
current in the non-linear response, via a mirror effect in the
continuum which introduces a resonance in the transmission
coefficient. However, in the non-linear response, the true continuum
is shielded from the bound state by the semi-continuum. This weakens
the resonance and suppresses the signatures of resonant transmission
in the heat current. The transmission coefficient is plotted in
Fig.~\ref{fig:jE_nonlinear}(c). For low transparency, the resonance
is weakened when $\Del T$ is increased. More specifically, the
area under the transmission curve, which is the relevant quantity
for the heat current, is reduced. For high transparency, the
opposite happens: the transmission coefficients is
less suppressed near the gap edge as $\Del T$ is increased (not
shown).

In addition to the spectral change found in $\D$, the change of the
thermal occupation factors for the two reservoirs influences the heat
transfer. We have checked that the main effect on the heat current is
due to the spectral change, but the change in the occupation factors
reduces the overall effect of a phase bias on the heat current. The
total heat transfer is larger for a larger temperature bias, but we
eliminate this scale factor by normalizing the heat current by the
corresponding heat current at zero phase bias, or at $T_c$.

The phase-modulation of the heat current is plotted in
Fig.~\ref{fig:jE_nonlinear}(b), while the temperature dependence
is shown in Fig.~\ref{fig:jE_nonlinear}(a). Since the angle
dependence of the barrier transparency does not qualitatively
affect our results, we use the angle-independent model. The
phase-modulation is weakened, and the peak effect near $T_c$ for
low transparency and phase difference $\phi=\pi$ is reduced.
However, the non-linearity does not remove or drastically
alter the resonance effects, and the main features
found in linear response persist for finite temperature bias.

\section{\label{sec:diffusive}Diffusive case}

In this section we discuss the case where the two leads are
dirty superconductors, i.e. the elastic mean free path, $\ell\ll
\xi_{\text{\tiny $\Delta$}}$. In dirty superconductors,
$\check{\G}(\mathbf{p}_f)$ is nearly
isotropic, and
\begin{equation}
\check{g}(\vR;\epsilon)={1\over -i\pi}\int {d\Omega_{\vp_f}\over
4\pi}\check{\G}(\vp_f,\vR;\epsilon,) =\left(\begin{array}{cc}
 \hat{g}^R & \hat{g}^K \\
0 & \hat{g}^A
\end{array} \right)
\,,
\end{equation}
satisfies the Usadel diffusion-type equation.\cite{usa70} The isotropic propagator
determines most of the physical observables. For example, the heat
current density is given by
\begin{equation}
\mathbf{j}_{\epsilon}={N_fv_f\ell \over 12}\int d\epsilon \,\epsilon\,
\mathrm{Tr}(\check{g}\grad\check{g})^K .
\label{hcd}
\end{equation}

To calculate the heat current through the point contact with dirty leads,
we need the boundary conditions for the isotropic propagator, $\check{g}$, at
an interface. Such boundary conditions were derived by Nazarov in the formulation
of the circuit theory of diffusive hetero-structures.\cite{naz99}
In this model the contact is described by a set of transmission eigenvalues,
$\{D_n\}$, or equivalently a distribution function, $\rho(D)$. The boundary
conditions are
\begin{equation}
N_{f1}v_{f1} \ell_1 \check{g}_1\partial_z\check{g}_1=
N_{f2}v_{f2} \ell_2 \check{g}_2\partial_z\check{g}_2
\,,
\label{dfbc1}
\end{equation}
\begin{equation}
{2\pi\over 3} \Area N_{f2}v_{f2}\ell_2
\check{g}_2\partial_z\check{g}_2=
\int dD \rho(D) \check{I}
\,,
\label{dfbc2}
\end{equation}
where the Keldysh matrix $\check{I}$ is defined as
\begin{equation}
\check{I}\equiv
{2D[\check{g}_2,\check{g}_1] \over 4+D(\{\check{g}_2,\check{g}_1\}-2)}
\,,
\end{equation}
$N_{fi}$ is the normal-state density of states at the Fermi level of
lead $i=1$, $2$, $v_{fi}$ is the Fermi velocity of lead $i$, and
$\Area$ is the cross-sectional area of the contact interface.
Eq.~\ref{dfbc1} is the conservation of spectral current, and
Eqs.~\ref{hcd}-\ref{dfbc2} imply that the total heat
current through the contact can be calculated from $\check{I}$,
\begin{equation}
\text{I}_{\epsilon}={1\over 8\pi}\int d\epsilon
\epsilon \int dD \rho(D)\,\mathrm{Tr}(\hat{I}^K)
\,.
\end{equation}
The Keldysh component of $\check{I}$ can be worked out to be\cite{naz99}
\begin{eqnarray*}
\hat{I}^K&=&{D\hat{g}^R_1\hat{g}^K_2\over 2-D+D\{\hat{g}^R_1,\hat{g}^R_2\}/2}+
{D\hat{g}^K_1\hat{g}^A_2\over 2-D+D\{\hat{g}^A_1,\hat{g}^A_2\}/2}\\
&-&{D^2(\hat{g}^R_1\hat{g}^K_2\hat{g}^A_1\hat{g}^A_2+\hat{g}^R_1\hat{g}^R_2\hat{g}^K_1\hat{g}^A_2
+2\hat{g}^K_2\hat{g}^A_2)
\over 2(2-D+D\{\hat{g}^R_1,\hat{g}^R_2\}/2)(2-D+D\{\hat{g}^A_1,\hat{g}^A_2\}/2)}\\
&-&(1\leftrightarrow 2) ,
\end{eqnarray*}
where $(1\leftrightarrow 2)$ means the exchange of index 1 and 2 in
all three terms.

For point contacts, we approximate $\hat{g}_{i}^{R,A,K}$ with
their bulk values, e.g.
$\hat{g}^K_i=(\hat{g}^R_i-\hat{g}^A_i)\tanh({\epsilon/2T_i})$. We
focus on the linear response, assume that superconducting
leads 1 and 2 have the same gap and evaluate the conductance in the limit
$\delta T\rightarrow 0$. After some algebra we find
\begin{eqnarray}
I_{\epsilon}=-{\delta T\over 2\pi T^2}
\int_{\Delta}^{\infty} d\epsilon \int dD \rho(D) \epsilon^2
(\epsilon^2-\Delta^2) \mathrm{sech}^2 {\epsilon\over 2T} \nonumber \\
\times {D(\epsilon^2-\Delta^2\cos{\phi\over 2})
+DR\Delta^2\sin^2(\phi/2)\over
[\epsilon^2-\Delta^2+D\Delta^2\sin^2(\phi/2)]^2}
\,.
\label{df4}
\end{eqnarray}
Notice that with the replacement,
\begin{equation}
\int dD \rho(D)(...)\Rightarrow 2\pi \Area N_fv_f\langle ...\rangle
\,,
\end{equation}
the result for the conductance with diffusive leads, Eq.~\ref{df4},
in the limit of a narrow distribution of transmission barriers,
is the exactly the same as Eqs.~\ref{eq:heat-linear}-\ref{eq:D-linear}
for the conductance obtained in the ballistic limit.

The near equivalence of the conductances for the diffusive and ballistic limits
is related in part to the small expansion parameter,
$a/\xi_{\text{\tiny$\Delta$}}$. In this limit pairbreaking effects resulting
from the current flow in the vicinity of the junction can be neglected, and as
a result the impurity renormalization of the excitation energy and order parameter
cancel for s-wave superconductors. The
resulting thermal conductance of point contact is then insensitive to
impurity scattering in the leads. This result is not expected to hold for
larger area contacts with $a\gtrsim\xi_{\text{\tiny$\Delta$}}$. These
junctions are beyond the scope of this work.

\section{Conclusions}\label{sec:conclusions}
We have derived a general expression for the phase- and temperature-dependent
thermal current through small Josephson weak links. The results are valid for
arbitrary transparency of the junction. In the ballistic limit, we obtained an
expression for the heat current in terms of a transmission coefficient,
$\D(\epsilon,\phi,T_1,T_2)$, for heat transport by continuum quasiparticle
states. The transmission coefficient includes direct transmission processes,
$\D_{ee}$, and transmission processes with branch conversion, $\D_{he}$. The
phase modulation, temperature dependence, and the dependence on the barrier
transparency of the heat current can be understood intuitively in terms of the
properties of the transmission coefficient.

In linear response, for high transparencies, the suppression of
the continuum density of states associated with the formation of a low-energy
Andreev bound state leads to a suppression of the
transmission coefficient, and the thermal conductance, as the phase
difference across the junction is tuned from $0$ to $\pi$. As a
consequence, the thermal conductance drops below the conductance for $\phi=0$
for temperatures below $T_c$.

However, for intermediate to low transparencies, a resonance develops in
the transmission coefficient as the phase difference is tuned from $0$
to $\pi$. Consequently, the conductance is larger at $\phi=\pi$ than at
$\phi=0$. The transmission resonance is due to the ABS being close to
the gap edge. This effect is similar to transmission resonances found in
wave mechanics for a quantum well with a shallow bound state just below
the continuum edge. For $\phi=\pi$ the resonance in the heat current
leads to an {\it increase} in the thermal conductance compared to the
normal-state conductance at $T_c$, over a broad temperature region below
$T_c$.

In the low-transparency limit, $D\ll 1$, we derived an analytic
expression for the heat conductance and found that there are
non-analytic corrections, of type $D\ln D$, to the usual linear in $D$
term. Also the phase dependence contains non-analytic terms of the
form $\sin^2\frac{\phi}{2}\ln(\sin^2\frac{\phi}{2})$. The
first-order tunnel Hamiltonian method gives a divergent result in
linear response, which is regulated by the formation of an Andreev bound
state from the continuum states near $\epsilon=\Delta$.
A similar divergence occurs in the sub-harmonic gap structure (SGS) of
the current-voltage characteristics of a superconducting tunnel
junction when the sub-gap structure is computed to finite order in
perturbation theory within the tunnel Hamiltonian method.\cite{sch63}
In the case of the SGS, summation to infinite order within a
wave-function method,\cite{bra95,ave95} or the tH method,\cite{cue96}
regulates the divergences. Such a summation is possible for the
thermal conductance within the tH method. However, the quasiclassical
Green's function method with the interface boundary conditions includes
the bound-state spectrum so unphysical divergences never appear.

We also studied the case when the superconducting leads
are in the diffusive limit. Based on the junction boundary conditions
developed by Nazarov, we found that the heat conductance has the same
form as in the ballistic case.

Finally, in the non-linear response we find that the resonance in the
transmission coefficient is reduced by the presence of the
semi-continuum spectrum, which shields the extended continuum energy
quasiparticle states carrying heat from the ABS. The effects
found in the linear response are reduced, but not dramatically.
Thus, we conclude that it is not essential to be in the linear response
limit in order to observe the resonance effects in the heat current.

\begin{acknowledgments}
This work was supported in part by the NSF grant DMR 9972087, and STINT, the
Swedish Foundation for International Cooperation in Research and
Higher Education. JAS acknowledges the support and hospitality of the
Aspen Center for Physics where this manuscript was completed.
\end{acknowledgments}

\appendix*
\section{Perturbation Theory}

We review the first-order perturbation calculation of the heat current for
low-transparency junctions based on the tunnelling Hamiltonian (tH)
method.\cite{amb63}
Consider two superconductors, labelled by $L$ and $R$, weakly coupled
by an insulating layer. They are assumed to be
described by the Hamiltonian
\[
H_{total}=H_0+H_T,
\]
where $H_0$ is the sum of the BCS reduced Hamiltonian of superconductors $L$ and $R$,
\[
H_0=H_L+H_R ,
\]
\[
H_L=\sum_{p,\sigma}\epsilon_p c^{\dagger}_{p,\sigma}c_{p,\sigma}+{1\over 2}\sum_{p,p',\sigma}V_{pp'}
c^{\dagger}_{p,\sigma}c^{\dagger}_{-p,-\sigma}c_{-p',-\sigma}c_{p',\sigma} ,
\]
\[
H_R=\sum_{k,\sigma}\epsilon_k a^{\dagger}_{k,\sigma}a_{k,\sigma}+{1\over 2}\sum_{k,k',\sigma}V_{kk'}
a^{\dagger}_{k,\sigma}a^{\dagger}_{-k,-\sigma}a_{-k',-\sigma}c_{k',\sigma} ,
\]
and $H_T$ describes the tunnelling processes,
\[
H_T=\sum_{k,p,\sigma}[\T_{kp}a{\dagger}_{k,\sigma}c_{p\sigma}+h.c.].
\]
The two superconductors are also assumed to be at the same chemical potential
$\mu_L=\mu_R$, but at different temperatures $T_L\neq T_R$.

The momentum index, $p$, and the annihilation operator,
$c_{p,\sigma}$, are reserved for superconductor $L$, while index, $k$, and
operator, $a_{k,\sigma}$, are reserved for superconductor $R$. The spin
state is labelled by $\sigma$, and $T_{kp}$ is the tunnelling matrix element. With
the BCS approximation, $V_{pp'}=\lambda < 0$, the order parameter is
defined as
\[ \Delta_L=-\lambda\sum_p\left< c_{-p,-\sigma}c_{p,\sigma}\right>=|\Delta_{L}|e^{i\phi_{L}} .
\]
The heat current operator is defined as
\[ I_{\epsilon}=-{\partial Q_L\over \partial t}=-i[H_{total},Q_L]=-i[H_T,Q_L] ,
\]
\[
Q_L=H_L-\mu_L \sum_{p,\sigma}c^{\dagger}_{p,\sigma}c_{p,\sigma} .
\]
It is straightforward to work out the commutator $[H_T,Q_L]$. Within
the mean field approximation, the pair annihilation operator
$c_{-q,-\sigma}c_{q,\sigma}$ reduces to its average value, and we find
\begin{equation}
I_{\epsilon}= -i\sum_{k,p,\sigma} \left[\T_{kp} \xi_p a^{\dagger}_{k,\sigma}c_{p,\sigma}
-\T_{kp}\Delta_L a^{\dagger}_{k,\sigma}c^{\dagger}_{-p,-\sigma}-h.c.\right] ,\label{hc-op}
\end{equation}
with $\xi_p=\epsilon_p-\mu_L$.

To calculate the ensemble average of the operator
$I_{\epsilon}$, we use the interaction picture and treat $H_T$
as a perturbation to $H_0$. According to first order perturbation
theory, the heat current has the form
\[
\text{I}_{\epsilon}=\langle I_{\epsilon}\rangle=-i\int_{-\infty}^t dt' e^{\eta t'}
\left<\left[\mathcal{I}_{\epsilon}(t),\mathcal{H}_T(t')\right]\right>_0 ,
\]
where $\mathcal{I}_{\epsilon}(t)$ and $\mathcal{H}_T(t')$ are the
operators in the interaction picture corresponding to
$I_{\epsilon}$ and $H_T$. respectively. The ensemble average
$\left<...\right>_0$ is defined by $H_0$, and $\eta=0^+$.
The commutator $[\mathcal{I}_{\epsilon}(t),\mathcal{H}_T(t')]$
contains various correlation functions,
\[
\text{I}_{\epsilon}=-\int_{-\infty}^{\infty} dt' e^{\eta t'} \Theta(t-t')
\sum_{i,j=1,2}[C_{ij}(t,t')+h.c.] ,
\]
where we define $C_{ij}(t,t')=\left<[\mathcal{O}_i(t),\mathcal{P}_j(t')]\right>_0$ with
\begin{eqnarray*}
&&\mathcal{O}_1(t)=\sum_{k,p,\sigma} \T_{kp} \xi_p a^{\dagger}_{k,\sigma}(t)c_{p,\sigma}(t) ,\\
&&\mathcal{O}_2(t)=-\sum_{k,p,\sigma} \T_{kp}\Delta_L a^{\dagger}_{k,\sigma}(t)c^{\dagger}_{-p,-\sigma}(t) ,\\
&&\mathcal{P}_1(t)=\sum_{k,p,\sigma} \T_{kp} a^{\dagger}_{k,\sigma}(t)c_{p,\sigma}(t),\\
&&\mathcal{P}_2(t)=\mathcal{P}_1^{\dagger}(t) ,
\end{eqnarray*}
and $a_{k,\sigma}(t)$ and $c_{p,\sigma}(t)$ are annihilation operators in the
interaction picture. We introduce the following single-particle correlation functions,
\begin{eqnarray*}
&&G^>_{p,\sigma}(t,t')=-i\left<c_{p,\sigma}(t)c^{\dagger}_{p,\sigma}(t')\right>_0 ,\\
&&G^<_{p,\sigma}(t,t')=+i\left<c^{\dagger}_{p,\sigma}(t')c_{p,\sigma}(t)\right>_0 ,\\
&&F^>_{p,\sigma}(t,t')=-i\left<c_{-p,-\sigma}(t)c_{p,\sigma}(t')\right>_0 ,\\
&&F^<_{p,\sigma}(t,t')=+i\left<c_{p,\sigma}(t')c_{-p,-\sigma}(t)\right>_0 ,\\
&&\bar{F}^>_{p,\sigma}(t,t')=-i\left<c^{\dagger}_{p,\sigma}(t)c^{\dagger}_{-p,-\sigma}(t')\right>_0 ,\\
&&\bar{F}^<_{p,\sigma}(t,t')=+i\left<c^{\dagger}_{-p,-\sigma}(t')c^{\dagger}_{p,\sigma}(t)\right>_0 .
\end{eqnarray*}
The correlation functions, $C_{ij}$, can then be decomposed into $G$'s
and $F$'s with the aid of Wick's theorem. For example,
\begin{eqnarray*}
&&C_{11}(t,t')=-\sum_{k,p,\sigma} \T_{kp}\T_{-k-p}\xi_p
\\
&&\times [\bar{F}^>_{k,\sigma}(t,t')F^<_{p,\sigma}(t',t)-
\bar{F}^<_{k,\sigma}(t,t')F^>_{p,\sigma}(t',t)] .
\end{eqnarray*}
The contribution of $C_{11}$ to the heat current is
\begin{eqnarray*}
\text{I}_{11}&=&-\int_{-\infty}^{\infty} dt' e^{\eta t'} \Theta(t-t')C_{11}(t,t') \\
&=&-2i\sum_{k,p} \T_{kp}\T_{-k-p}\xi_p \int {d\omega \over 2\pi}\int {d\omega' \over 2\pi} \\
&&[f_L(\omega)-f_R(\omega')]{B(p,\omega)\bar{B}(k,\omega')
\over \omega'-\omega-i\eta} .
\end{eqnarray*}
The factor 2 comes from the sum over spin, and $A$ and $B$ are
spectral functions, with
\begin{eqnarray*}
&&G^{\gtrless}_p(\omega)=\mp i f_L(\mp\omega)A(p,\omega) ,\\
&&F^{\gtrless}_p(\omega)=\mp i f_L(\mp\omega)B(p,\omega) ,\\
&&\bar{F}^{\gtrless}_p(\omega)=\mp i f_L(\mp\omega)\bar{B}(p,\omega) .
\end{eqnarray*}
Here we introduced the Fermi function
$f_L(\omega)=(e^{\omega/T_L}+1)^{-1}$, and the spectral functions
\begin{eqnarray*}
&&A(p,\omega)=\pi[(1+{\xi_p\over E_p})\delta(\omega-E_p)+(1-{\xi_p\over E_p})\delta(\omega+E_p)] ,\\
&&B(p,\omega)=\pi{\Delta_L\over E_p}[\delta(\omega-E_p)-\delta(\omega+E_p)] ,\\
&&\bar{B}(p,\omega)=B^*(p,\omega),
\end{eqnarray*}
with $E_p=(\xi_p^2+\Delta_L^2)^{1/2}$. Analogous calculations
for the other correlation functions follow. Summing up
all the contributions for the heat current gives
\begin{eqnarray}
&&\text{I}_{\epsilon}=-4\,\mathrm{Im}\sum_{k,p}\int {d\omega \over 2\pi}
\int {d\omega' \over 2\pi}[f_L(\omega)-f_R(\omega')] \times
\nonumber \\
&&\Big\{ |\T_{kp}|^2\xi_p {A(p,\omega)A(k,\omega')
\over \omega'-\omega+i\eta}
+|\T_{kp}|^2\Delta_L {\bar{B}(-p,\omega)A(k,\omega')\over
 \omega'-\omega+i\eta} \nonumber \\
&&-\T_{kp}\T_{-k-p}\xi_p{B(p,\omega)\bar{B}(k,\omega')
\over \omega'-\omega-i\eta} \nonumber \\
&&+T_{kp}\T_{-k-p}\Delta_L{A(-p,\omega)\bar{B}(k,\omega')
\over \omega'-\omega-i\eta}\Big\} .\label{deno}
\end{eqnarray}

In Eq.~\ref{deno} we neglect the momentum dependence of the
tunnelling matrix elements,  put $\T_{kp}\rightarrow {\bf T}$,
and change the momentum sum into integrals over excitation energies
$E_p$ and $E_k$. The integrals over $\omega$ and $\omega'$ collapse
because of the $\delta$ functions in the spectral functions $A$, $B$
and $\bar{B}$. Taking the imaginary part of the integrand yields new
$\delta$ functions $\delta(E_k\pm E_p)$, which lead to a collapse of
one of the energy integrals. Finally, the heat current takes the form
\begin{eqnarray}
I_{\epsilon}=8\pi N_f^L N_f^R |{\bf T}|^2
\int^{+\infty}_{\Delta_{\mathrm{max}}}d\epsilon \;\epsilon
[ f_L(\epsilon)-f_R(\epsilon) ] \nonumber \\
\times
{\epsilon^2-|\Delta_L||\Delta_R| \cos\phi \over
\sqrt{\epsilon^2-|\Delta_L|^2} \sqrt{\epsilon^2-|\Delta_R|^2}} .\label{int}
\end{eqnarray}
Here the phase difference, $\phi=\phi_{L}-\phi_{R}$,
$\Delta_{\mathrm{max}}=\mathrm{max}[|\Delta_{L}|,|\Delta_{R}|]$, and
$N_f^{L/R}$ are the density of states at the Fermi level for $L$ and
$R$ sides, respectively.

Finally, we compare with previous calculations of the heat current in
tunnel junctions. In Ref.~\onlinecite{mak65} only the kinetic
energy term was included in deriving the current operator; the term,
$T_{kp}\Delta_La^{\dagger}_{k,\sigma}c^{\dagger}_{-p,-\sigma}$, and its
Hermitian conjugate were missing from Eq.~\ref{hc-op}. In the main
result of Guttman et al. (Eq.~9 of Ref.~\cite{gut97}) the sign of the
$\cos\phi$ term is incorrect.

-----------------------------------------------------------------------

\begin{thebibliography}{31}
\expandafter\ifx\csname natexlab\endcsname\relax\def\natexlab#1{#1}\fi
\expandafter\ifx\csname bibnamefont\endcsname\relax
  \def\bibnamefont#1{#1}\fi
\expandafter\ifx\csname bibfnamefont\endcsname\relax
  \def\bibfnamefont#1{#1}\fi
\expandafter\ifx\csname citenamefont\endcsname\relax
  \def\citenamefont#1{#1}\fi
\expandafter\ifx\csname url\endcsname\relax
  \def\url#1{\texttt{#1}}\fi
\expandafter\ifx\csname urlprefix\endcsname\relax\def\urlprefix{URL }\fi
\providecommand{\bibinfo}[2]{#2}
\providecommand{\eprint}[2][]{\url{#2}}

\bibitem[{\citenamefont{Belzig et~al.}(1999)\citenamefont{Belzig, Wilhelm,
  Bruder, {Sch\"on}, and Zaikin}}]{bel99}
\bibinfo{author}{\bibfnamefont{W.}~\bibnamefont{Belzig}},
  \bibinfo{author}{\bibfnamefont{F.~K.} \bibnamefont{Wilhelm}},
  \bibinfo{author}{\bibfnamefont{C.}~\bibnamefont{Bruder}},
  \bibinfo{author}{\bibfnamefont{G.}~\bibnamefont{{Sch\"on}}},
  \bibnamefont{and} \bibinfo{author}{\bibfnamefont{A.}~\bibnamefont{Zaikin}},
  \bibinfo{journal}{Superlatt. Microstruct.} \textbf{\bibinfo{volume}{25}},
  \bibinfo{pages}{1251} (\bibinfo{year}{1999}).

\bibitem[{\citenamefont{Agrait et~al.}(2003)\citenamefont{Agrait, Yeyati, and
  van Ruitenbeek}}]{agr03}
\bibinfo{author}{\bibfnamefont{N.}~\bibnamefont{Agrait}},
  \bibinfo{author}{\bibfnamefont{A.~L.} \bibnamefont{Yeyati}},
  \bibnamefont{and} \bibinfo{author}{\bibfnamefont{J.~M.} \bibnamefont{van
  Ruitenbeek}}, \bibinfo{journal}{Phys. Rep.} \textbf{\bibinfo{volume}{377}},
  \bibinfo{pages}{81} (\bibinfo{year}{2003}).

\bibitem[{\citenamefont{Aumentado et~al.}(1999)\citenamefont{Aumentado,
  Chandrasekhar, Eom, Baldo, and Rehn}}]{aum99}
\bibinfo{author}{\bibfnamefont{J.}~\bibnamefont{Aumentado}},
  \bibinfo{author}{\bibfnamefont{V.}~\bibnamefont{Chandrasekhar}},
  \bibinfo{author}{\bibfnamefont{J.}~\bibnamefont{Eom}},
  \bibinfo{author}{\bibfnamefont{P.~M.} \bibnamefont{Baldo}}, \bibnamefont{and}
  \bibinfo{author}{\bibfnamefont{L.~E.} \bibnamefont{Rehn}},
  \bibinfo{journal}{Appl. Phys. Lett.} \textbf{\bibinfo{volume}{75}},
  \bibinfo{pages}{3554} (\bibinfo{year}{1999}).

\bibitem[{\citenamefont{Zhao et~al.}(2003)\citenamefont{Zhao, L\"ofwander, and
  Sauls}}]{zha03c}
\bibinfo{author}{\bibfnamefont{E.}~\bibnamefont{Zhao}},
  \bibinfo{author}{\bibfnamefont{T.}~\bibnamefont{L\"ofwander}},
  \bibnamefont{and} \bibinfo{author}{\bibfnamefont{J.~A.} \bibnamefont{Sauls}},
  \bibinfo{journal}{arXiv/cond-mat/} \textbf{\bibinfo{volume}{0302346}},
  \bibinfo{pages}{4} (\bibinfo{year}{2003}),
  \bibinfo{note}{[to appear in PRL, 2003]}.

\bibitem[{\citenamefont{Eschrig et~al.}(1999)\citenamefont{Eschrig, Sauls, and
  Rainer}}]{esc99}
\bibinfo{author}{\bibfnamefont{M.}~\bibnamefont{Eschrig}},
  \bibinfo{author}{\bibfnamefont{J.~A.} \bibnamefont{Sauls}}, \bibnamefont{and}
  \bibinfo{author}{\bibfnamefont{D.}~\bibnamefont{Rainer}},
  \bibinfo{journal}{Phys. Rev. B} \textbf{\bibinfo{volume}{60}},
  \bibinfo{pages}{10447} (\bibinfo{year}{1999}).

\bibitem[{\citenamefont{Eschrig}(2000)}]{esc00}
\bibinfo{author}{\bibfnamefont{M.}~\bibnamefont{Eschrig}},
  \bibinfo{journal}{Phys. Rev. B} \textbf{\bibinfo{volume}{61}},
  \bibinfo{pages}{9061} (\bibinfo{year}{2000}).

\bibitem[{\citenamefont{Josephson}(1962)}]{jos62}
\bibinfo{author}{\bibfnamefont{B.~D.} \bibnamefont{Josephson}},
  \bibinfo{journal}{Phys. Lett.} \textbf{\bibinfo{volume}{1}},
  \bibinfo{pages}{251} (\bibinfo{year}{1962}).

\bibitem[{\citenamefont{Arnold}(1987)}]{arn87}
\bibinfo{author}{\bibfnamefont{G.~B.} \bibnamefont{Arnold}},
  \bibinfo{journal}{J. Low Temp. Phys.} \textbf{\bibinfo{volume}{68}},
  \bibinfo{pages}{1} (\bibinfo{year}{1987}).

\bibitem[{\citenamefont{Andreev}(1965)}]{and65}
\bibinfo{author}{\bibfnamefont{A.~F.} \bibnamefont{Andreev}},
  \bibinfo{journal}{Zh. Eksp. Teor. Fiz.} \textbf{\bibinfo{volume}{49}},
  \bibinfo{pages}{655} (\bibinfo{year}{1965}), \bibinfo{note}{[Sov Phys JETP
  {\bf 22} 455]}.

\bibitem[{\citenamefont{Andreev}(1964)}]{and64}
\bibinfo{author}{\bibfnamefont{A.~F.} \bibnamefont{Andreev}},
  \bibinfo{journal}{Zh. Eksp. Teor. Fiz.} \textbf{\bibinfo{volume}{46}},
  \bibinfo{pages}{1823} (\bibinfo{year}{1964}), \bibinfo{note}{[Sov Phys JETP
  {\bf 19} 1228 (1964)]}.

\bibitem[{\citenamefont{Maki and Griffin}(1965)}]{mak65}
\bibinfo{author}{\bibfnamefont{K.}~\bibnamefont{Maki}} \bibnamefont{and}
  \bibinfo{author}{\bibfnamefont{A.}~\bibnamefont{Griffin}},
  \bibinfo{journal}{Phys. Rev. Lett.} \textbf{\bibinfo{volume}{15}},
  \bibinfo{pages}{921} (\bibinfo{year}{1965}).

\bibitem[{\citenamefont{Guttman et~al.}(1997)\citenamefont{Guttman, Nathanson,
  Ben-Jacob, and Bergman}}]{gut97}
\bibinfo{author}{\bibfnamefont{G.~D.} \bibnamefont{Guttman}},
  \bibinfo{author}{\bibfnamefont{B.}~\bibnamefont{Nathanson}},
  \bibinfo{author}{\bibfnamefont{E.}~\bibnamefont{Ben-Jacob}},
  \bibnamefont{and} \bibinfo{author}{\bibfnamefont{D.~J.}
  \bibnamefont{Bergman}}, \bibinfo{journal}{Phys. Rev. B}
  \textbf{\bibinfo{volume}{55}}, \bibinfo{pages}{3849} (\bibinfo{year}{1997}).

\bibitem[{\citenamefont{Guttman et~al.}(1998)\citenamefont{Guttman,
  Ben-Jacob, and Bergman}}]{gut98}
\bibinfo{author}{\bibfnamefont{G.~D.} \bibnamefont{Guttman}},
  \bibinfo{author}{\bibfnamefont{E.}~\bibnamefont{Ben-Jacob}},
  \bibnamefont{and} \bibinfo{author}{\bibfnamefont{D.~J.}
  \bibnamefont{Bergman}}, \bibinfo{journal}{Phys. Rev. B}
  \textbf{\bibinfo{volume}{57}}, \bibinfo{pages}{2717} (\bibinfo{year}{1998}).

\bibitem[{\citenamefont{Kulik and Ome\'{l}yanchuk}(1978)}]{kul78}
\bibinfo{author}{\bibfnamefont{I.}~\bibnamefont{Kulik}} \bibnamefont{and}
  \bibinfo{author}{\bibfnamefont{A.}~\bibnamefont{Ome\'{l}yanchuk}},
  \bibinfo{journal}{Sov. J. Low Temp. Phys.} \textbf{\bibinfo{volume}{4}},
  \bibinfo{pages}{142} (\bibinfo{year}{1978}).

\bibitem[{\citenamefont{Eliashberg}(1972)}]{eli72}
\bibinfo{author}{\bibfnamefont{G.~M.} \bibnamefont{Eliashberg}},
  \bibinfo{journal}{Zh. Eskp. Teor. Fiz.} \textbf{\bibinfo{volume}{61}},
  \bibinfo{pages}{1254} (\bibinfo{year}{1972}), \bibinfo{note}{[English
  translation Sov. Phys. JETP {\bf 34}, 668 (1972)]}.

\bibitem[{\citenamefont{Larkin and Ovchinnikov}(1975)}]{lar76}
\bibinfo{author}{\bibfnamefont{A.}~\bibnamefont{Larkin}} \bibnamefont{and}
  \bibinfo{author}{\bibfnamefont{Y.}~\bibnamefont{Ovchinnikov}},
  \bibinfo{journal}{Zh. Eskp. Teor. Fiz.} \textbf{\bibinfo{volume}{68}},
  \bibinfo{pages}{1915} (\bibinfo{year}{1975}), \bibinfo{note}{[English
  translation Sov. Phys. JETP {\bf 41}, 960 (1976)]}.

\bibitem[{\citenamefont{Larkin and Ovchinnikov}(1977)}]{lar77}
\bibinfo{author}{\bibfnamefont{A.}~\bibnamefont{Larkin}} \bibnamefont{and}
  \bibinfo{author}{\bibfnamefont{Y.}~\bibnamefont{Ovchinnikov}},
  \bibinfo{journal}{Zh. Eskp. Teor. Fiz.} \textbf{\bibinfo{volume}{73}},
  \bibinfo{pages}{299} (\bibinfo{year}{1977}), \bibinfo{note}{[English
  translation Sov. Phys. JETP {\bf 46}, 155 (1978)]}.

\bibitem[{\citenamefont{Rammer and Smith}(1986)}]{ram86}
\bibinfo{author}{\bibfnamefont{J.}~\bibnamefont{Rammer}} \bibnamefont{and}
  \bibinfo{author}{\bibfnamefont{H.}~\bibnamefont{Smith}},
  \bibinfo{journal}{Rev. Mod. Phys.} \textbf{\bibinfo{volume}{58}},
  \bibinfo{pages}{323} (\bibinfo{year}{1986}).

\bibitem[{\citenamefont{Serene and Rainer}(1983)}]{ser83}
\bibinfo{author}{\bibfnamefont{J.~W.} \bibnamefont{Serene}} \bibnamefont{and}
  \bibinfo{author}{\bibfnamefont{D.}~\bibnamefont{Rainer}},
  \bibinfo{journal}{Phys. Rep.} \textbf{\bibinfo{volume}{101}},
  \bibinfo{pages}{221} (\bibinfo{year}{1983}).

\bibitem[{\citenamefont{Nagato et~al.}(1993)\citenamefont{Nagato, Nagai, and
  Hara}}]{nag93}
\bibinfo{author}{\bibfnamefont{Y.}~\bibnamefont{Nagato}},
  \bibinfo{author}{\bibfnamefont{K.}~\bibnamefont{Nagai}}, \bibnamefont{and}
  \bibinfo{author}{\bibfnamefont{J.}~\bibnamefont{Hara}}, \bibinfo{journal}{J.
  Low Temp. Phys.} \textbf{\bibinfo{volume}{93}}, \bibinfo{pages}{33}
  (\bibinfo{year}{1993}).

\bibitem[{\citenamefont{Schopohl and Maki}(1995)}]{sch95}
\bibinfo{author}{\bibfnamefont{N.}~\bibnamefont{Schopohl}} \bibnamefont{and}
  \bibinfo{author}{\bibfnamefont{K.}~\bibnamefont{Maki}},
  \bibinfo{journal}{Physica} \textbf{\bibinfo{volume}{B 204}},
  \bibinfo{pages}{214} (\bibinfo{year}{1995}).

\bibitem[{\citenamefont{Za\u{i}tsev}(1984)}]{zai84}
\bibinfo{author}{\bibfnamefont{A.~V.} \bibnamefont{Za\u{i}tsev}},
  \bibinfo{journal}{Sov. Phys. JETP} \textbf{\bibinfo{volume}{59}},
  \bibinfo{pages}{1015} (\bibinfo{year}{1984}).

\bibitem[{\citenamefont{{L\"ofwander} et~al.}(2003)\citenamefont{{L\"ofwander},
  {Fogelstr\"om}, and Sauls}}]{lof03c}
\bibinfo{author}{\bibfnamefont{T.}~\bibnamefont{{L\"ofwander}}},
  \bibinfo{author}{\bibfnamefont{M.}~\bibnamefont{{Fogelstr\"om}}},
  \bibnamefont{and} \bibinfo{author}{\bibfnamefont{J.~A.} \bibnamefont{Sauls}},
  \bibinfo{journal}{arXiv/cond-mat/} \textbf{\bibinfo{volume}{0304588}},
  \bibinfo{pages}{15} (\bibinfo{year}{2003}),
  \bibinfo{note}{[to appear in PRB, 2003]}.

\bibitem[{\citenamefont{Kulik and Ome\'{l}yanchuk}(1992)}]{kul92}
\bibinfo{author}{\bibfnamefont{I.}~\bibnamefont{Kulik}} \bibnamefont{and}
  \bibinfo{author}{\bibfnamefont{A.}~\bibnamefont{Ome\'{l}yanchuk}},
  \bibinfo{journal}{Sov. J. Low Temp. Phys.} \textbf{\bibinfo{volume}{18}},
  \bibinfo{pages}{819} (\bibinfo{year}{1992}).

\bibitem[{\citenamefont{Usadel}(1970)}]{usa70}
\bibinfo{author}{\bibfnamefont{K.}~\bibnamefont{Usadel}},
  \bibinfo{journal}{Phys. Rev. Lett.} \textbf{\bibinfo{volume}{25}},
  \bibinfo{pages}{507} (\bibinfo{year}{1970}).

\bibitem[{\citenamefont{Nazarov}(1999)}]{naz99}
\bibinfo{author}{\bibfnamefont{Y.~V.} \bibnamefont{Nazarov}},
  \bibinfo{journal}{Superlatt. Microstruc.} \textbf{\bibinfo{volume}{25}},
  \bibinfo{pages}{1221} (\bibinfo{year}{1999}),
  \bibinfo{note}{[also in cond-mat/9811155]}.

\bibitem[{\citenamefont{Schrieffer and Wilkins}(1963)}]{sch63}
\bibinfo{author}{\bibfnamefont{J.~R.} \bibnamefont{Schrieffer}}
  \bibnamefont{and} \bibinfo{author}{\bibfnamefont{J.~W.}
  \bibnamefont{Wilkins}}, \bibinfo{journal}{Phys. Rev. Lett.}
  \textbf{\bibinfo{volume}{10}}, \bibinfo{pages}{17} (\bibinfo{year}{1963}).

\bibitem[{\citenamefont{Bratus et~al.}(1995)\citenamefont{Bratus, Shumeiko, and
  Wendin}}]{bra95}
\bibinfo{author}{\bibfnamefont{E.~N.} \bibnamefont{Bratus}},
  \bibinfo{author}{\bibfnamefont{V.~S.} \bibnamefont{Shumeiko}},
  \bibnamefont{and} \bibinfo{author}{\bibfnamefont{G.}~\bibnamefont{Wendin}},
  \bibinfo{journal}{Phys. Rev. Lett.} \textbf{\bibinfo{volume}{74}},
  \bibinfo{pages}{2110} (\bibinfo{year}{1995}).

\bibitem[{\citenamefont{Averin and Bardas}(1995)}]{ave95}
\bibinfo{author}{\bibfnamefont{D.}~\bibnamefont{Averin}} \bibnamefont{and}
  \bibinfo{author}{\bibfnamefont{A.}~\bibnamefont{Bardas}},
  \bibinfo{journal}{Phys. Rev. Lett.} \textbf{\bibinfo{volume}{75}},
  \bibinfo{pages}{1831} (\bibinfo{year}{1995}).

\bibitem[{\citenamefont{Cuevas et~al.}(1996)\citenamefont{Cuevas,
  Martin-Rodero, and Yeyati}}]{cue96}
\bibinfo{author}{\bibfnamefont{J.~C.} \bibnamefont{Cuevas}},
  \bibinfo{author}{\bibfnamefont{A.}~\bibnamefont{Martin-Rodero}},
  \bibnamefont{and} \bibinfo{author}{\bibfnamefont{A.~L.}
  \bibnamefont{Yeyati}}, \bibinfo{journal}{Phys. Rev. B}
  \textbf{\bibinfo{volume}{54}}, \bibinfo{pages}{7366} (\bibinfo{year}{1996}).

\bibitem[{\citenamefont{Ambegaokar and Baratoff}(1963)}]{amb63}
\bibinfo{author}{\bibfnamefont{V.}~\bibnamefont{Ambegaokar}} \bibnamefont{and}
  \bibinfo{author}{\bibfnamefont{A.}~\bibnamefont{Baratoff}},
  \bibinfo{journal}{Phys. Rev. Lett.} \textbf{\bibinfo{volume}{10}},
  \bibinfo{pages}{486} (\bibinfo{year}{1963}).

\end{thebibliography}
-----------------------------------------------------------------------

\end{document}